\documentclass[sigconf]{acmart}

\renewcommand\footnotetextcopyrightpermission[1]{} % removes footnote with conference information in first column
\usepackage{color}
\usepackage{graphicx}
\usepackage{algorithm2e}
\usepackage{epsfig}
\usepackage{amssymb}
\usepackage{multicol}
\usepackage{subfigure}
\usepackage[bottom]{footmisc}
\usepackage{verbatim}
\usepackage{enumitem}
\usepackage{xcolor}
\usepackage{colortbl}
\usepackage{makecell}

\DeclareMathOperator*{\argmin}{\emph{arg\,min}}

\begin{document}
\title{RAPPER: {\color{blue}Ra}nsomware {\color{blue}P}revention via {\color{blue}Per}formance Counters}

%\author{, Sarani Bhattacharya, Debdeep Mukhopadhyay, Anupam Chattopadhyay}
\author{Manaar Alam}
%\authornote{Dr.~Trovato insisted his name be first.}
%\orcid{1234-5678-9012}
\affiliation{%
  \institution{Indian Institute of Technology Kharagpur}
  %\streetaddress{P.O. Box 1212}
  %\city{Kharagpur}
  %\state{Ohio}
  %\postcode{43017-6221}
}
\email{alam.manaar@iitkgp.ac.in}

\author{Sarani Bhattacharya}
%\authornote{Dr.~Trovato insisted his name be first.}
%\orcid{1234-5678-9012}
\affiliation{%
  \institution{Indian Institute of Technology Kharagpur}
  %\streetaddress{P.O. Box 1212}
  %\city{Kharagpur}
  %\state{Ohio}
  %\postcode{43017-6221}
}
\email{sarani.bhattacharya@cse.iitkgp.ernet.in}

\author{Debdeep Mukhopadhyay}
%\authornote{Dr.~Trovato insisted his name be first.}
%\orcid{1234-5678-9012}
\affiliation{%
  \institution{Indian Institute of Technology Kharagpur}
  %\streetaddress{P.O. Box 1212}
  %\city{Kharagpur}
  %\state{Ohio}
  %\postcode{43017-6221}
}
\email{debdeep@cse.iitkgp.ernet.in}

\author{Anupam Chattopadhyay}
%\authornote{Dr.~Trovato insisted his name be first.}
%\orcid{1234-5678-9012}
\affiliation{%
  \institution{Nanyang Technological University Singapore}
  %\streetaddress{P.O. Box 1212}
  %\city{Kharagpur}
  %\state{Ohio}
  %\postcode{43017-6221}
}
\email{anupam@ntu.edu.sg}

\begin{abstract}
Ransomware can produce direct and controllable economic loss, which makes it one of the 
most prominent threats in cyber security. As per the latest statistics, more than half 
of malwares reported in Q1 of 2017 are ransomware and there is a potent threat of a 
novice cybercriminals accessing rasomware-as-a-service. 
The concept of public-key based data kidnapping and subsequent 
extortion was introduced in 1996. Since then, variants of ransomware 
emerged with different cryptosystems and larger key sizes though, the underlying techniques remained same.
Though there are works in literature which proposes a generic framework to detect 
the crypto ransomwares, we present a two step unsupervised detection tool which when suspects a process activity to be malicious, issues an alarm for further analysis to be carried in the second step and detects it with minimal traces.
The two step detection framework- RAPPER uses Artificial Neural Network and Fast Fourier Transformation to develop a highly accurate, fast and reliable solution to ransomware detection using minimal trace points.
%In this paper we use the event counter statistics from the HPCs to construct a detection framework RAPPER which monitors the anomaly in the behavior of these performance count values sampled over time. 
\end{abstract}
%\vspace{-0.3cm}

\maketitle  % typeset the title of the contribution
%\vspace{-0.2cm}
% Figure with Ransomware process 
\section{Introduction}
If your organization has not been hit by ransomware yet, there are chances that it will soon be. The number of medium to large-scale business falling prey to ransom payment and extortion of their private databases have increased manifold. 
These malicious executables infect the victim machine and demands a ransom amount after encrypting the files and documents of the machine. In May 2017, WannaCry ransomware has affected approx. 200000 business across 150 countries. 
Identification, blocking of these ransomwares at the earliest along with recovering the contents of the already encrypted files is already an open challenge.

Hardware Performance Counters (HPCs) were first introduced for checking the static and dynamic integrity of programs, for the purpose of detecting any malicious modifications to them as discussed in~\cite{malone2011hardware}. 
While in~\cite{demme2013feasibility} performance counters are used to build a malware detector in hardware. 
Detecting malware which modifies the kernel control flow has been targeted in ~\cite{karri:13, wang2016reusing}. The paper uses performance counters to monitor the system calls to detect the vulnerability. 
However, detection of ransomwares through the HPCs, to the best of our knowledge, has not been attempted so far. Though the underlying technique is similar~\cite{simha:14}, ransomware detection requires far more accuracy and faster response time to limit the damage.

A range of ransomwares were studied in~\cite{kharraz_2015_dimva}, which identified 15 different ransomware families. It is suggested that despite advancing encryption systems, the prominent ransomwares leave a trait in the access of IO and file-systems. Accordingly, Kharraz $et al.$~\cite{kharraz_2016_usenix_automate} proposed a technique of correlating high file system activity with the intrusion of ransomware, which, however, is susceptible to false positives and also can be defeated with a slow encryption process. Moreover, the technique requires modification in OS kernel, which may not be practical in many real scenarios.
In a recent work, Kiraz $et al.$ presented a technique, where large integer multiplication blocks are identified within an execution~\cite{kiraz_2017_eprint_arith}. Since public-key cryptosystems rely on large integer multiplications, it can detect the threat at an early stage. Similar approaches, for detection of symmetric-key cryptographic primitives via data flow graph isomorphism~\cite{Lestringant_2015_asiaccs_isomorph} or by identifying characteristics of a cipher in a binary code~\cite{grobert_2011_book_binary}, are also presented.
In this paper, neither we target a specific family of ransomwares nor the properties corresponding to a particular cipher implementation. Instead we develop a generic anomaly based approach based on the HPC statistics.
% discuss shortcomings
%\vspace{-0.3cm}
\subsection*{Motivation and Contribution}
The primary contributions of this paper are listed below:
\begin{itemize}{\small\vspace*{-0.1cm}
\item The main objective of RAPPER is to learn the behavior of the system under observation with performance event statistics obtained from HPCs. Unlike other works in literature, which save the templates of malicious processes and matches it on its occurrence, here we allow our tool to learn the normal operating behavior of the system. The time-series data as observed from a selected cluster of HPC events is fed to an Artificial Neural Network to learn the specific characteristics of the data.
\item Any deviation from this normal behavior as learned by the Autoencoder is considered as a suspect to RAPPER. Thus we observe that the performance statistics of the system in presence of ransomware are significantly dissimilar from the normal system behavior because of repeated encryption process.
\item In the course of our study, we identified a benign benchmark application which can raise a false alarm. This is typically a benign process, but because of its high computational overhead the system behavior differ significantly from its normal behavior thus raising a false alarm. 
\item Thus in the final step, we transform the time series to frequency domain using 
Fast Fourier Transformation (FFT) and understand the repeatability of data with help of a second autoencoder. 

}
\end{itemize}
RAPPER is a lightweight tool, which neither requires any root privilege, nor requires any hardware and kernel modification, thereby making it practical to use in almost every environment.

%\vspace{-0.2cm}
\section{Anomaly Detection by Analysing the System Behavior}
In this section, we first analyze the normal behavior of a system by monitoring some
appropriately selected hardware performance counters in parallel. We then present a
notion of anomalous activity in the system and demonstrate a detailed methodology
for detecting those anomalies by using an \texttt{Autoencoder}.
%\vspace{-0.2cm}

\begin{figure*}[!t]
	\centering
	\small %\vspace{-0.55cm}
	\subfigure[\textbf{\small \# Branch Instructions}]{
		\includegraphics[width=0.28\textwidth]{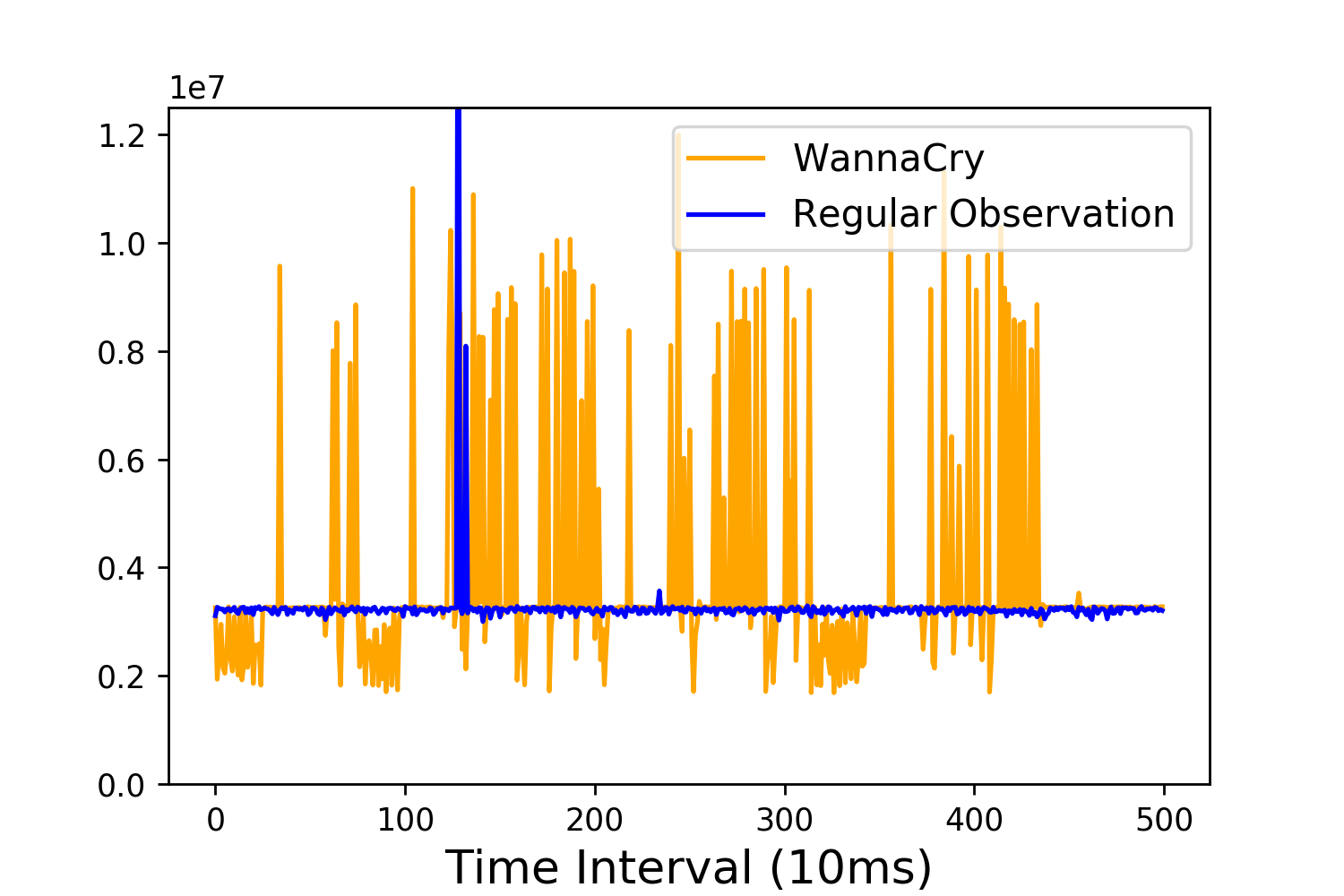}
		\label{fig:observe_branch}}
	~~
	\subfigure[\textbf{\small \# Branch Mispredictions observed}]{
		\includegraphics[width=0.28\textwidth]{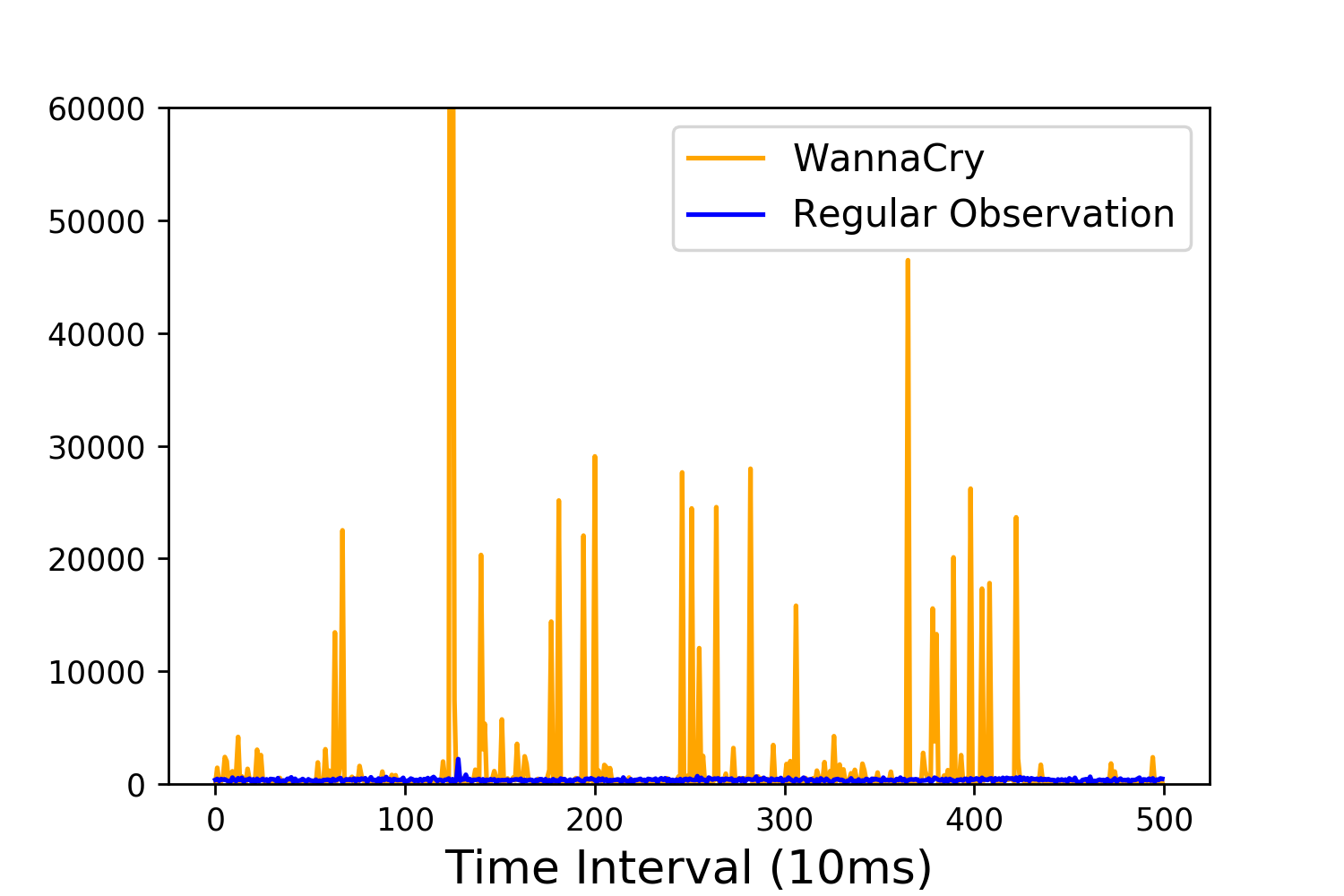}
		\label{fig:observe_br_miss}}%\vspace{-0.65cm}
	~~
	\subfigure[\textbf{\small \# Cache Misses encountered}]{
		\includegraphics[width=0.28\textwidth]{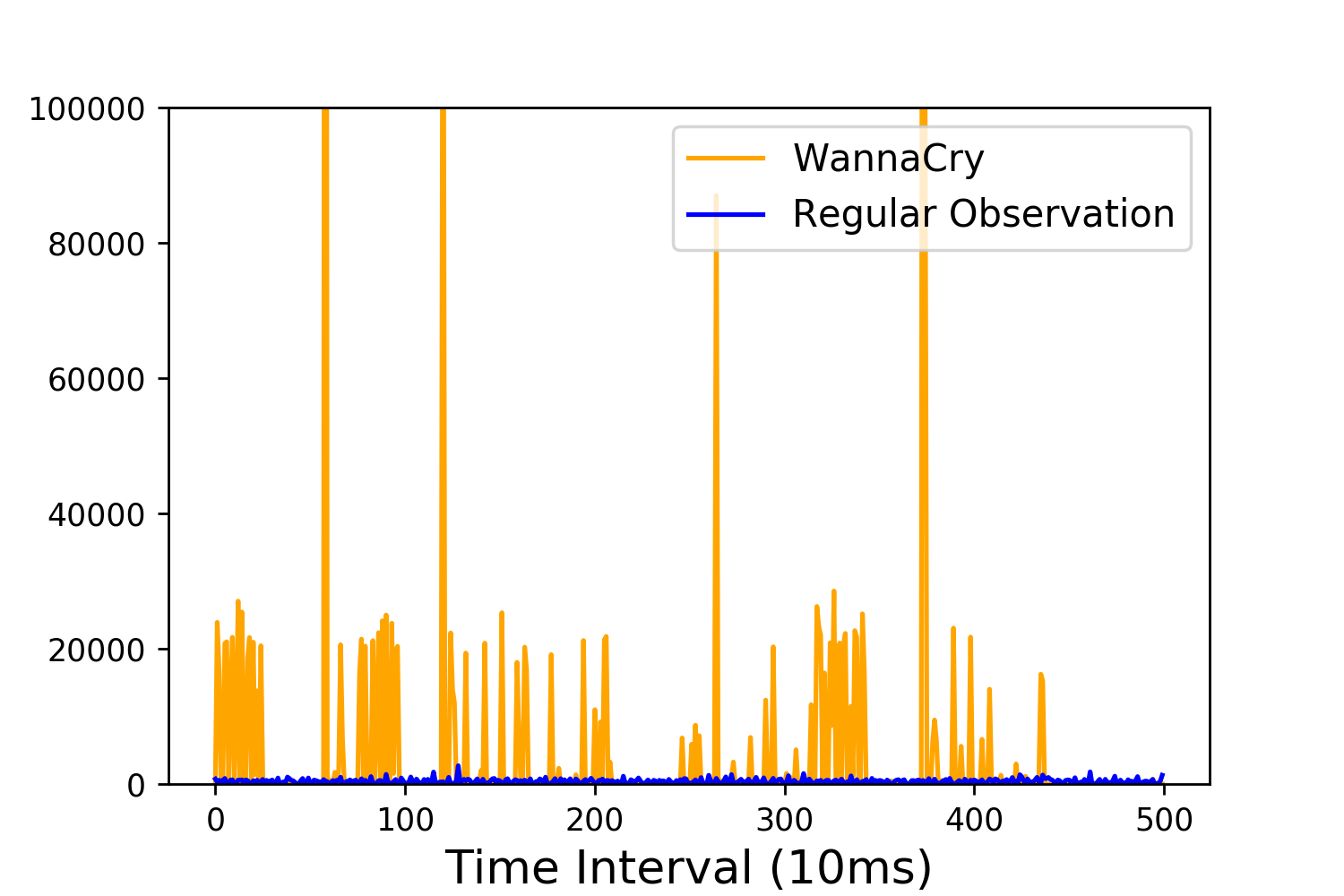}
		\label{fig:observe_ca_miss}}%\vspace{-0.65cm}
	\quad
	\subfigure[\textbf{\small \# Cache References observed }]{
		\includegraphics[width=0.28\textwidth]{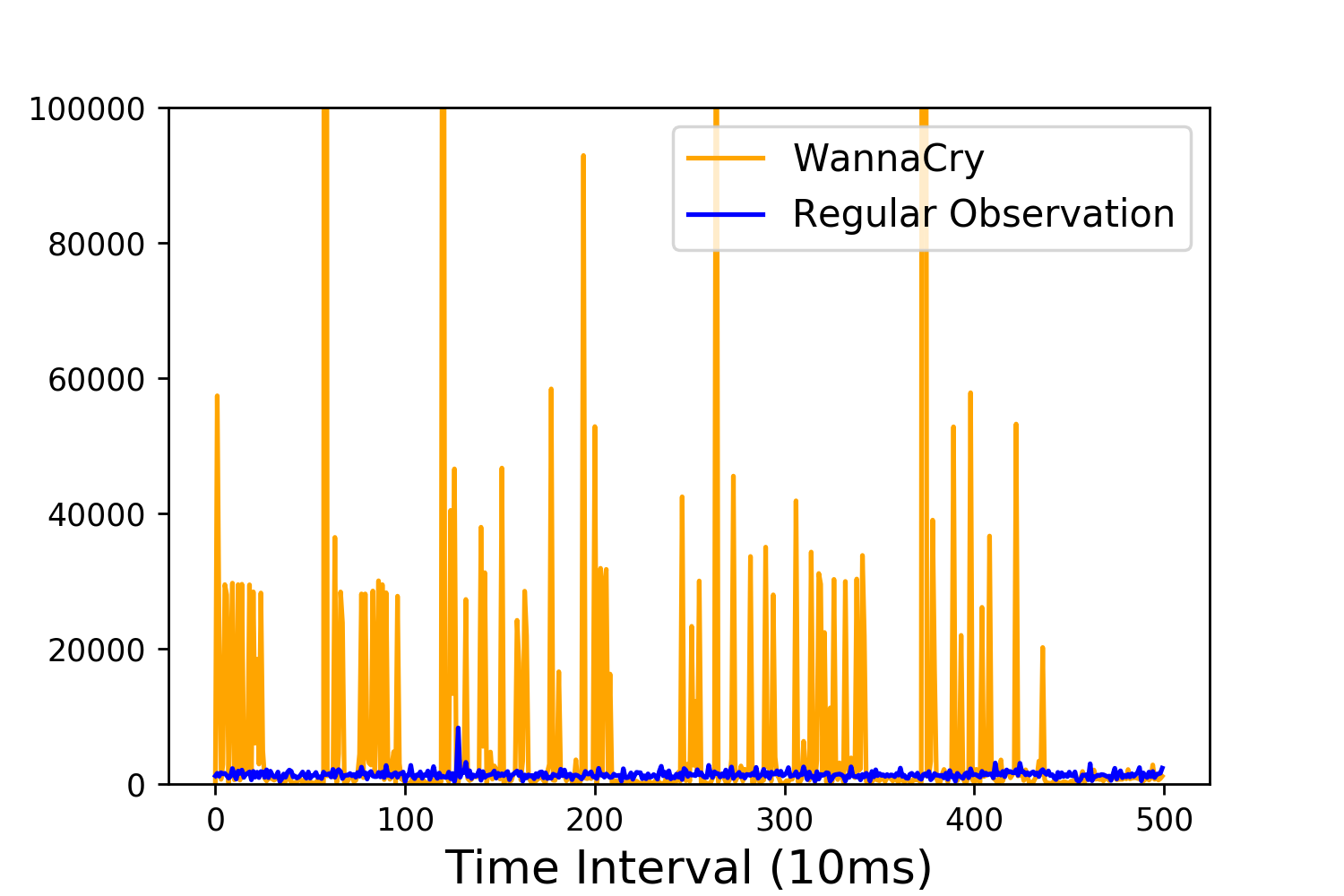}
		\label{fig:observe_cache}}%\vspace{-0.65cm}
	~~
	\subfigure[\textbf{\small \# Instructions observed}]{
		\includegraphics[width=0.28\textwidth]{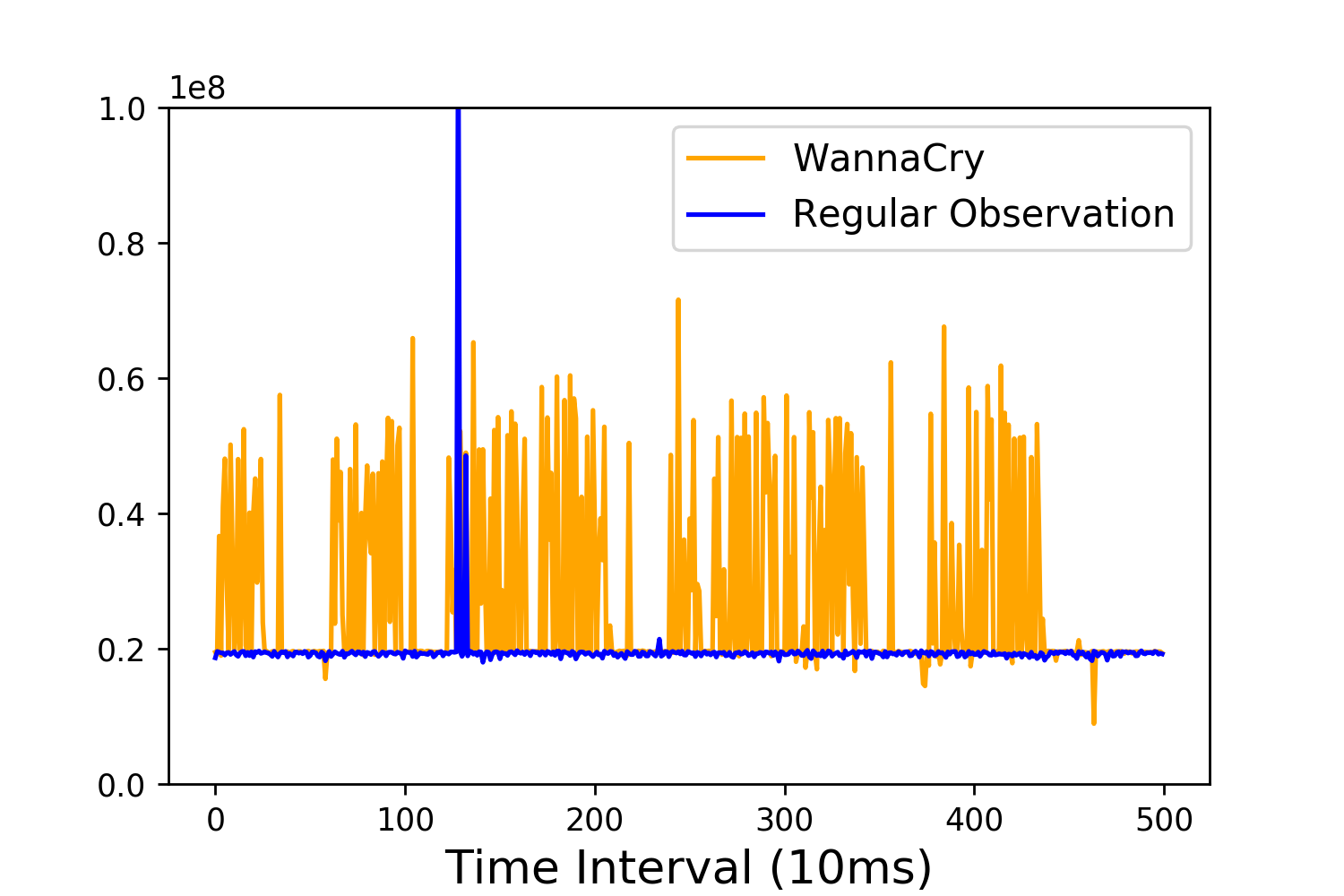}
		\label{fig:observe_ins}}\vspace*{-0.5cm}
	\caption{\small Variation of Performance Event Counters from HPCs 
		in presence of Wannacry Ransomware\label{fig:perf_dev}}%\vspace*{-0.4cm}
\end{figure*}
\subsection{Observing the system behavior using HPCs}\label{sec:hpc}
The Hardware Performance Counters (HPCs) are a set of special purpose registers built
into modern processors to dynamically observe the hardware related activities in a computer
system. There are some recent works~\cite{karri:13} which use these HPCs to detect
malicious programs targeted for a particular system. The HPCs can be monitored dynamically
with a user privilege using the well-known \texttt{perf} tool, available in Linux kernels 2.6.31+.
One interesting property of the perf tool is that a user can observe the performance counters
associated with a system with some time interval, thereby giving the benefit of observing the
system behavior continuously in a succession of time. The command to monitor a particular 
HPC event in such way is as follows:

\begin{center}
	\footnotesize \texttt{perf stat -e <event\_name> -I <time\_interval> <executable\_name>}
\end{center}

Since our objective is to detect the presence of ransomware, which mainly contains an encryption program, typically involving both symmetric and asymmetric key encryptions, we selected the hardware events which are more likely to change because of the
encryptions. The hardware events selected for our study are \texttt{instruction},
\texttt{cache-references}, \texttt{cache-misses}, \texttt{branches}, and \texttt{branch-misses}. The events are self-explanatory by their names.
Generally the symmetric encryption affects the cache based events while the asymmetric encryptions affect the instruction and branching events.

To represent the prototype of a typical system behavior, we designed a watchdog program, and 
collected the \texttt{perf stat} values with \texttt{10ms} time interval for that executable.
The effects of all the other processes including the ransomware running in the system will have an impact on the performance
counters. We collected these values at the different point of time in the target system and created
a dataset of regular observation. We articulate that any behavior which is not close to this dataset
is an unusual activity, but may not be a malicious one.

We show the effect of a Ransomware Program (for example, a WannaCry) on the HPCs in Figure~\ref{fig:perf_dev}.
The \textcolor{blue}{blue} lines in Figure represent the effect of normal system programs on the
watchdog executable for different HPCs, whereas the \textcolor{orange}{orange} lines show the effect
of WannaCry ransomware. 

An important point to be observed from Figure~\ref{fig:observe_br_miss} is that for a particular time interval the behavior
of WannaCry does not change much from the typical system behavior, for example around time
interval of 100 the effect of WannaCry on the hardware event branch misses is same as normal
system behavior. So, instead of considering individual points for decision making, we select a
window of observations considering each of the five events collectively. Thus, we transform
the problem into anomaly detection in multivariate time-series data.

%\vspace{-0.3cm}
\subsection{Learning a Time-Series data using an Autoencoder}\label{sec:learn_auto}
To present a generalized ransomware detection strategy we do not model
the behavior of the ransomwares as there can be potential new ransomware
whose behavior is unknown and cannot be modeled. Instead, we model the
normal system behavior, as we can get a majority of such instances. Another
advantage of detecting anomalies by modeling normal behavior is that we do
not need the necessity of labeled dataset as any activity with unusual behavior
crossing a certain threshold can be detected as an anomaly. Thus, we propose
an unsupervised approach to detect these anomalies. The HPC event counts observed over the watchdog application can be considered as the time-series data which is system dependent.
An LSTM (Long-Short-Term-Memory)
based autoencoder can efficiently implement the unsupervised anomaly
detection for time-series, which we discuss below.

Autoencoder is an Artificial Neural Network used for efficient coding of the
input space by unsupervised learning. The primary goal of an autoencoder
is to induce a representation for a set of data by learning an approximate
identity function, i.e., if the input data is $\mathcal{X}$, the goal of the 
autoencoder is to learn the function $f$, given by - \hspace{1cm}
%\begin{equation*}
	$f : \mathcal{X} \rightarrow \mathcal{X}$
%\end{equation*}

An autoencoder always consists of two mapping, encoding and decoding,
which are given as $\phi$ and $\psi$ respectively.
\begin{gather*}
	\phi : \mathcal{X} \rightarrow \mathcal{F}, \hspace{1cm}
	\psi : \mathcal{F} \rightarrow \mathcal{X}
\end{gather*}
where $\mathcal{F}$ is a vector referring to the decisive intermediate representation learned by the autoencoder,
which is used to regenerate the original input data. The error incurred by the autoencoder
to regenerate the input from vector $\mathcal{F}$ is termed as \texttt{Reconstruction Error},
which is given as below:
\begin{equation}
\mathcal{L} = \| \mathcal{X} - (\psi \circ \phi)\mathcal{X}\|^2
\end{equation}
The learning goal of the autoencoder is to minimize these reconstruction cost for all the input
samples, i.e., to find the mappings $\phi$ and $\psi$ such that $\mathcal{L}$ is minimum.
\begin{equation}
	\label{eq:rec_err}
	\argmin_{\phi, \psi} \mathcal{L} = \argmin_{\phi, \psi} \| \mathcal{X} - (\psi \circ \phi)\mathcal{X}\|^2
\end{equation}
In our case, the input $\mathcal{X}$ is a multivariate time-series sequence, and the
objective is to learn the structure of the sequence. LSTM networks, belonging to a 
class of Recurrent Neural Network Model, are typically
used for modeling sequence data, which efficiently handles the dependencies within
the sequence. Hence, we use the LSTM based autoencoder for our detection purpose.
The anomaly detector model first takes a multivariate input sequence ($\mathcal{X}$),
generates an intermediate feature vector ($\mathcal{F}$) related to the sequence,
and then reconstructs the same sequence from the intermediate feature vector. The
autoencoder is trained using all the input sequences by following the objective function
mentioned in Equation~(\ref{eq:rec_err}).

The training dataset is constructed from the observed data for normal system behavior 
by taking a window of 100 trace points (i.e., a window trace points collected over 1 second, since each interval
data is collected after 10ms).
Without loss of generality, we have chosen 100 trace points for our experiments.
We shift the window by one time-interval (i.e., 10ms) repeatedly to consider consecutive 100 sample point for learning.
Once the learning of intermediate vector $\mathcal{F}$ is completed, for an anomalous sequence, the autoencoder makes an attempt to reconstruct the original input sequence. Thus, the autoencoder maps it to the normal sequence, based on the intermediate vector $\mathcal{F}$. There is an inherent information loss in this process and hence will incur a substantial \texttt{reconstruction error}. Next, we quantify the amount of error to be incurred by
a process to be termed as an anomaly.
%\vspace{-0.2cm}
\subsection{Determining Threshold for Decision}\label{sec:threshold}
To quantify the threshold for detecting anomalous activities, we calculate the reconstruction
error distribution ($\mathcal{R}$) for all the training samples. According to the $3\sigma$ rule of thumb,
all the values are considered within three standard deviations of the mean.
Hence, we set the threshold for reconstruction error ($\mathcal{R}_t$) as below.
\begin{equation}
\label{eq:three_sigma}
\mathcal{R}_{t} = \mu_{\mathcal{R}} + 3 * \sigma_{\mathcal{R}}
\end{equation}
where $\mu_{\mathcal{R}}$ and $\sigma_{\mathcal{R}}$ are the mean and standard deviation of distribution $\mathcal{R}$.
In our experimental setup $\mathcal{R}_{t}$ came out to be $0.114$.
\vspace{-0.2cm}
\subsubsection{Anomalous Behaviors of Ransomwares}
\begin{figure}[!t]
		\centering
		\small %\vspace{-0.55cm}
		\subfigure[\textbf{\small WannaCry}]{
			\includegraphics[width=0.46\linewidth]{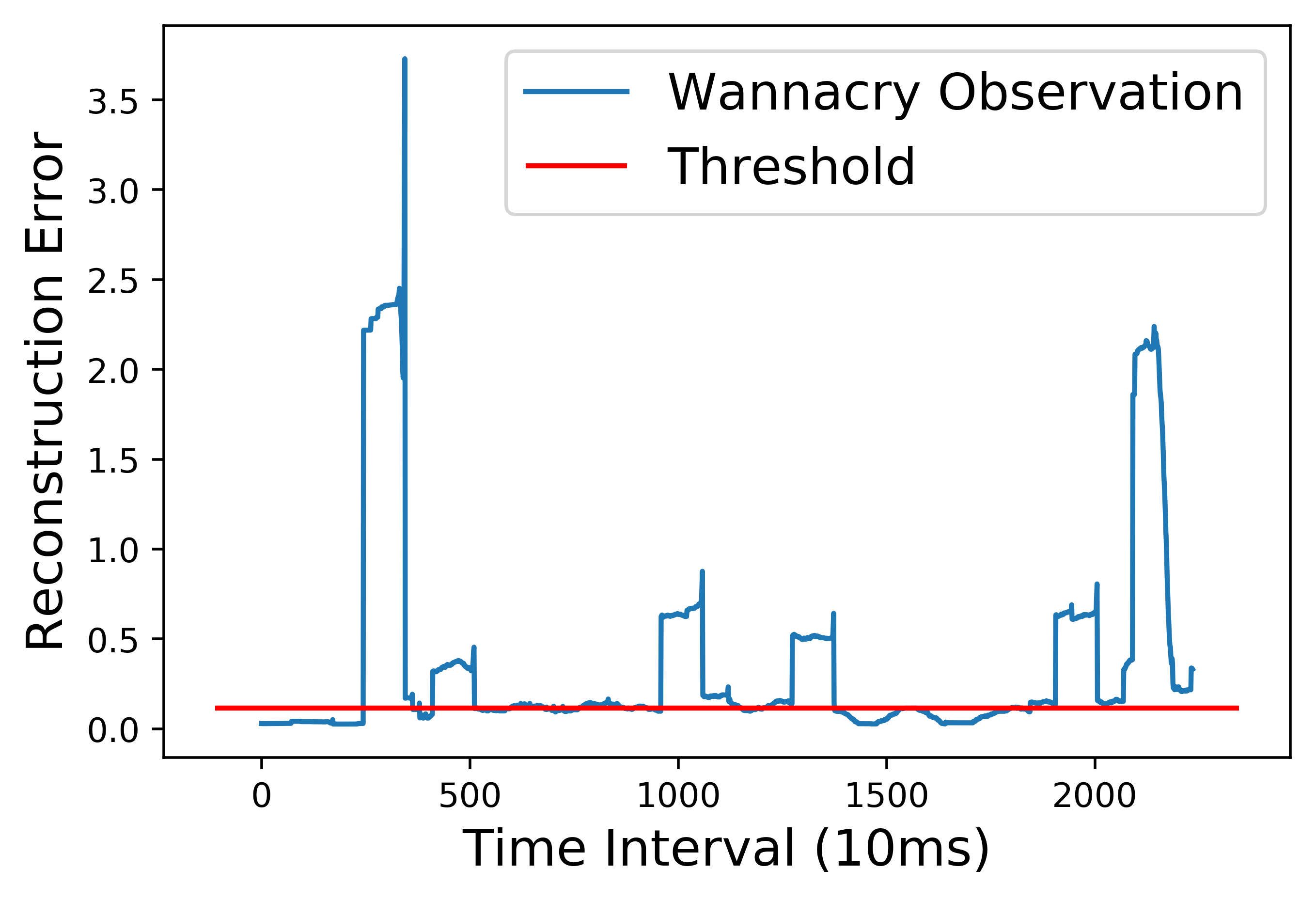}
			\label{fig:rec_wannacry}}
		\subfigure[\textbf{\small Vipasana}]{
			\includegraphics[width=0.46\linewidth]{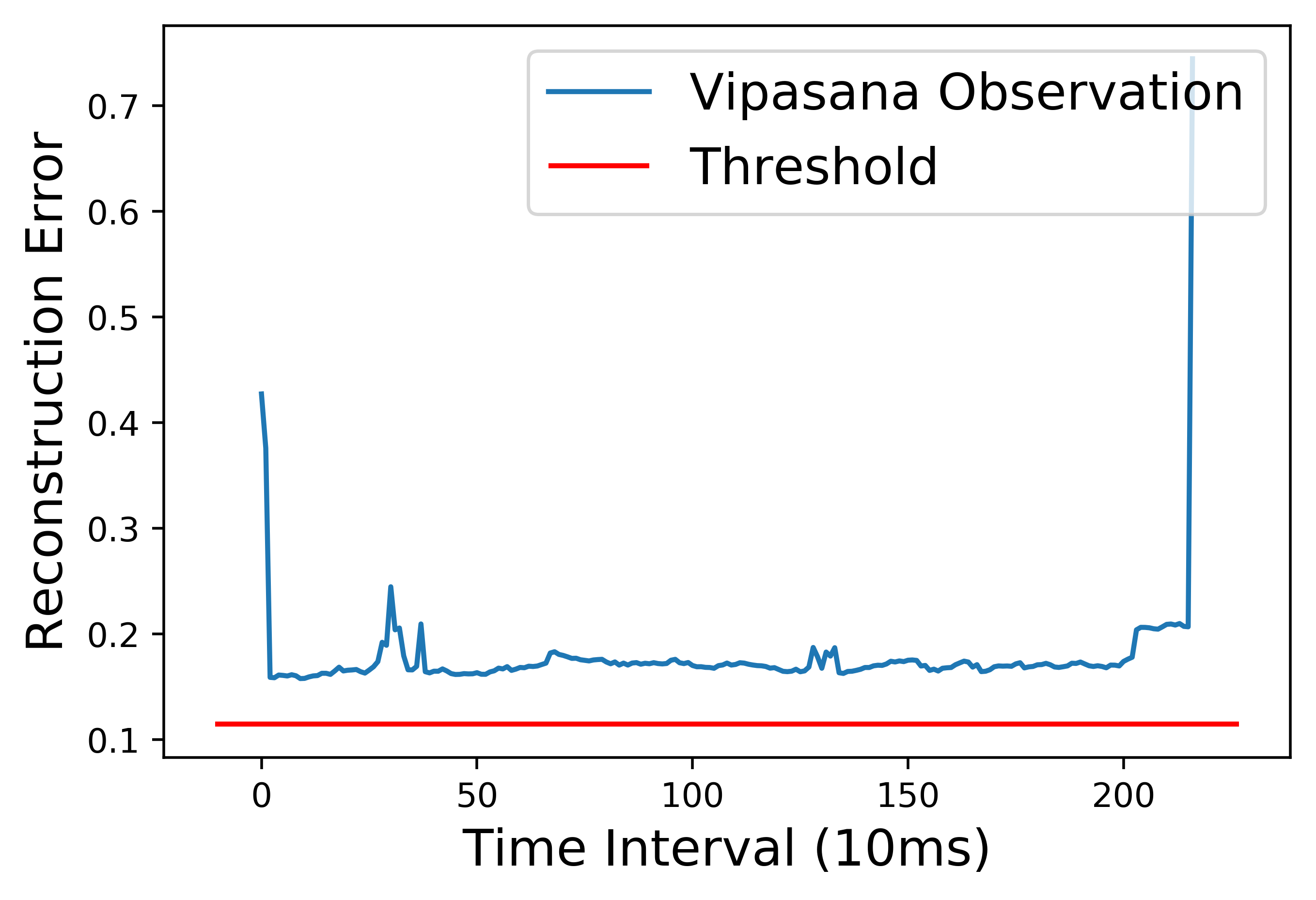}
			\label{fig:rec_vipasana}}\vspace*{-0.5cm}%\vspace{-0.65cm}
		\caption{\small Sequence of Reconstruction Errors for Ransomware\label{fig:seq_rec_error}}\vspace*{-0.6cm}
	\end{figure}

	In our study, we considered two ransomware programs - namely \emph{WannaCry} and \emph{Vipasana}
	to show the impact of selecting the threshold $\mathcal{R}_t$ in detecting them as anomalies.
	Figure~\ref{fig:seq_rec_error} shows the sequence of reconstruction errors for both the ransomwares.
	The first point on both the plot appears after observing the first window of 100 time-interval (equivalently
	1 second after the start of execution of the ransomwares). The successive points come after each
	time-interval of 10ms since we are sliding the window by one time-interval for calculating the next
	reconstruction error. The \textcolor{blue}{blue} line indicates the reconstruction errors of each window
	whereas the \textcolor{red}{red} line signifies the threshold $\mathcal{R}_t$ as calculated before.
	
	We can observe from Figure~\ref{fig:rec_wannacry}, the execution of WannaCry starts behaving like a
	regular program (since the reconstruction error lies well below the threshold value), but the reconstruction
	error shoots over the threshold at $245^{th}$ observation. Thus, the WannaCry
	is detected as anomaly $(1000 + 244*10) = 3440$ ms or $3.44$ seconds after the start of execution.
	Whereas, from Figure~\ref{fig:rec_vipasana}, we can observe that the Vipasana is detected as an anomaly
	at the first window itself, i.e., $1$ second after the start of execution. In both the cases there is an extra 
    overhead of time due to the testing time of Autoencoder, which we discuss in Section~\ref{sec:results}.
	%\newpage
	
% \newpage
%\vspace{-0.4cm}
\section{How good is Reconstruction error as a decider?}
In the previous section, we suggest that a threshold as high as $\mathcal{R}_t$ 
can be used to determine whether a particular application 
behavior deviates from the normal system behavior significantly. 
In this section, we explain why a single decision step is not enough 
to claim that the anomaly observed is from a malicious process.
%\vspace{-0.2cm}
	\subsection{Understanding the Ambiguity}
\begin{figure}[!t]
	\centering
	\small %\vspace{-0.55cm}
	\subfigure[\textbf{\small \# Branch Instructions}]{
		\includegraphics[width=0.2\textwidth]{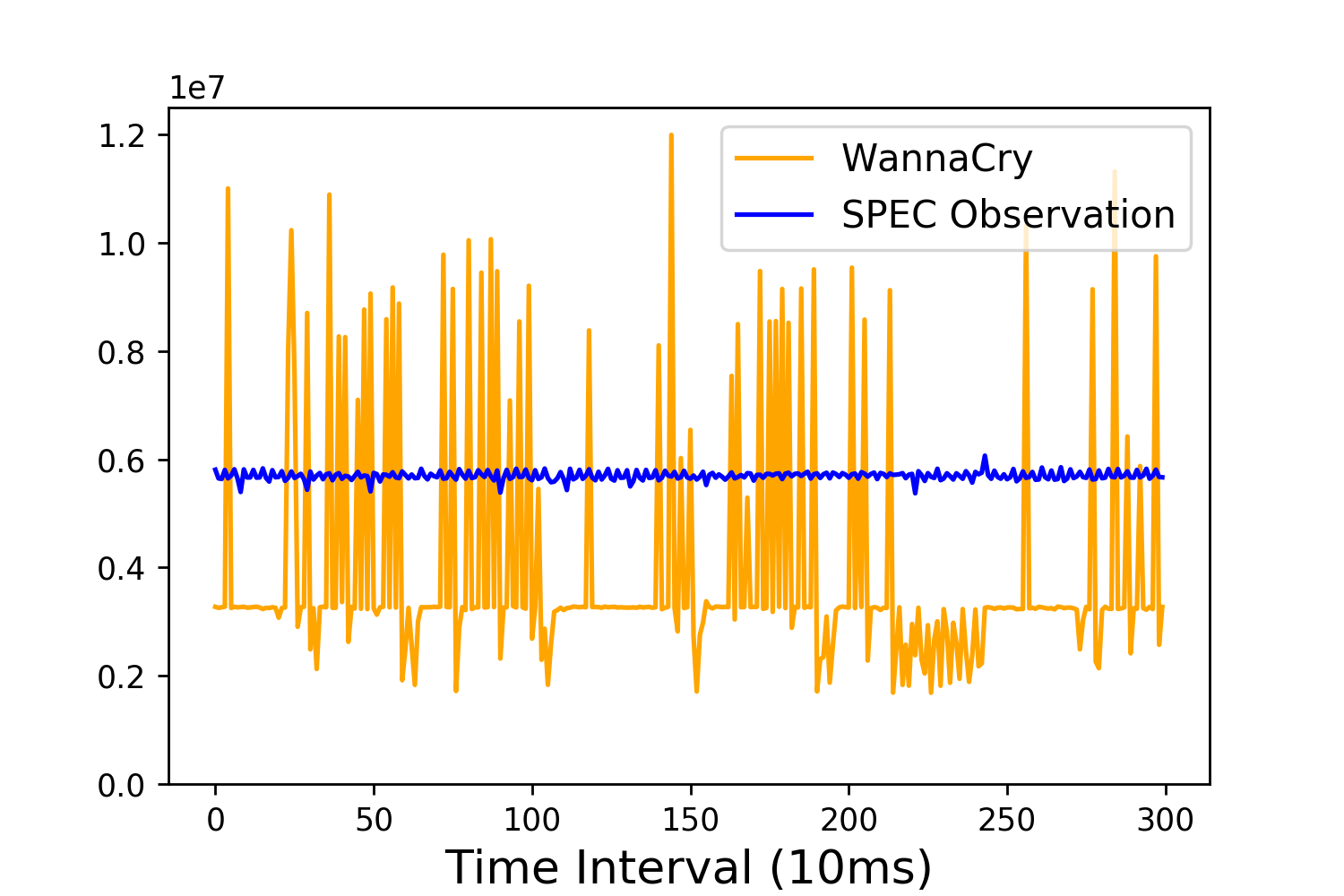}
		\label{fig:com_branch}}
	\quad
	\subfigure[\textbf{\small \# Branch Mispredictions}]{
		\includegraphics[width=0.2\textwidth]{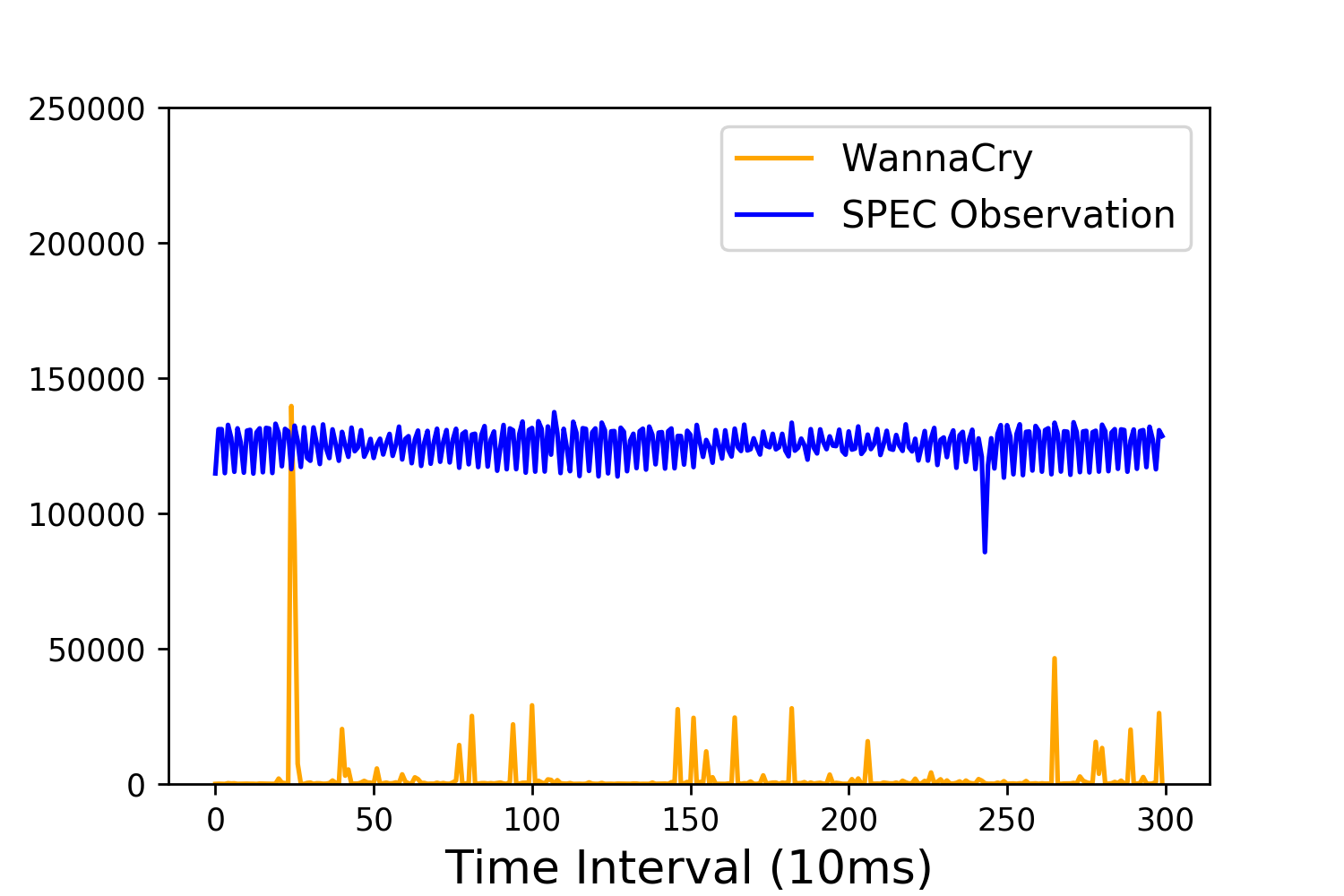}
		\label{fig:com_br_miss}}%\vspace{-0.65cm}
	\quad
	\subfigure[\textbf{\small \# cache misses}]{
		\includegraphics[width=0.2\textwidth]{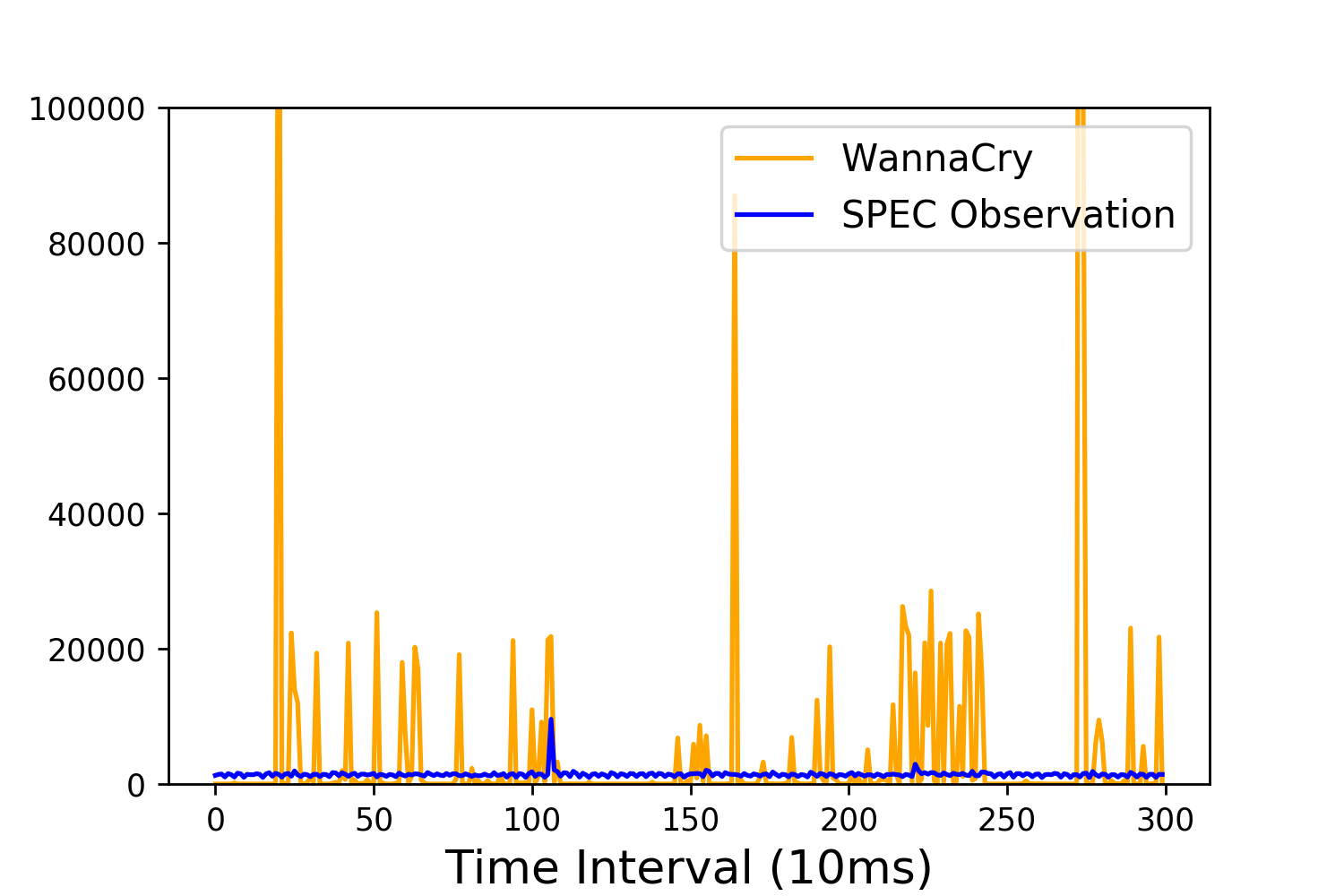}
		\label{fig:com_ca_miss}}%\vspace{-0.65cm}
	\quad
	\subfigure[\textbf{\small \# Cache References}]{
		\includegraphics[width=0.2\textwidth]{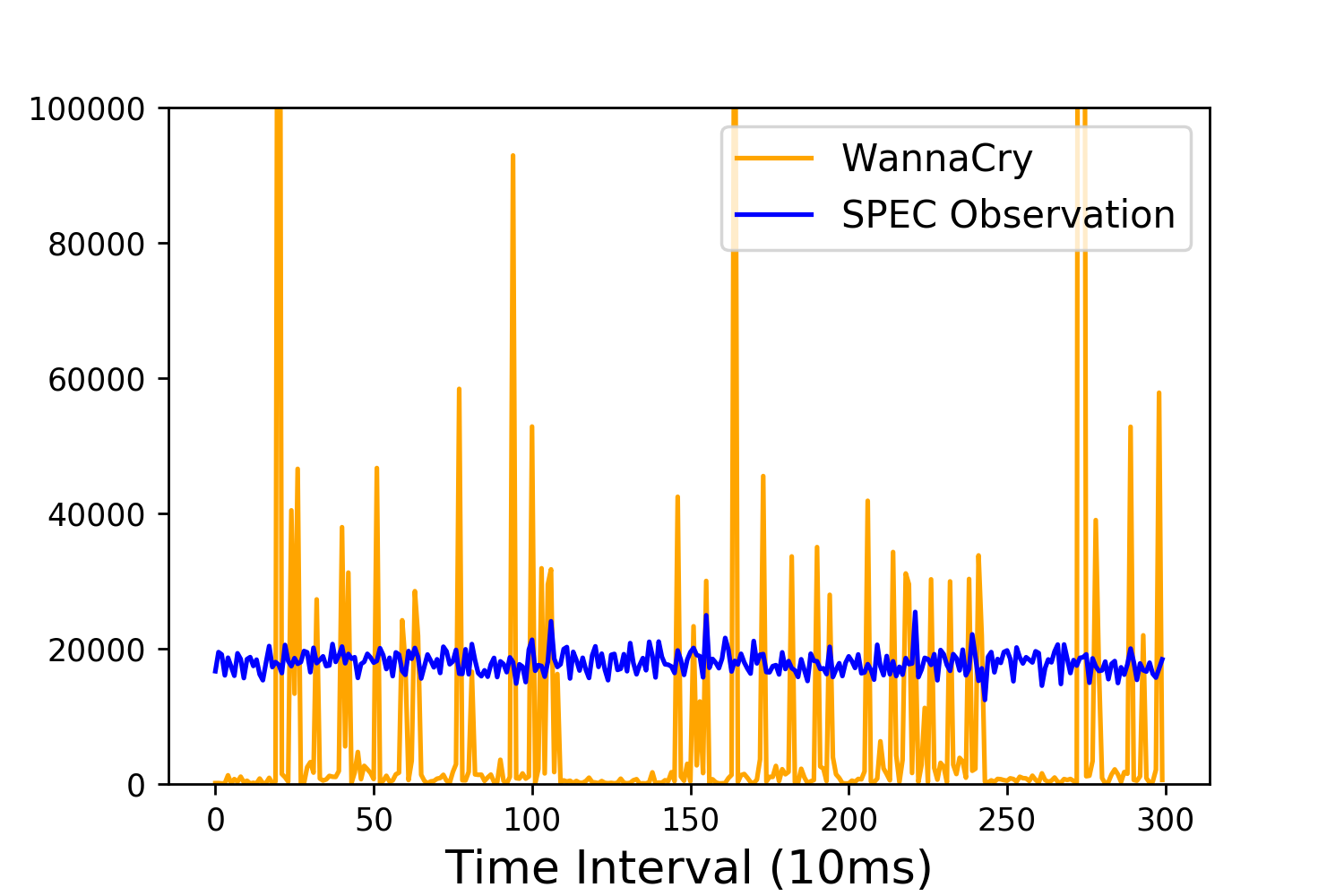}
		\label{fig:com_cache}}%\vspace{-0.65cm}
	\quad
	\subfigure[\textbf{\small \# Instructions}]{
		\includegraphics[width=0.2\textwidth]{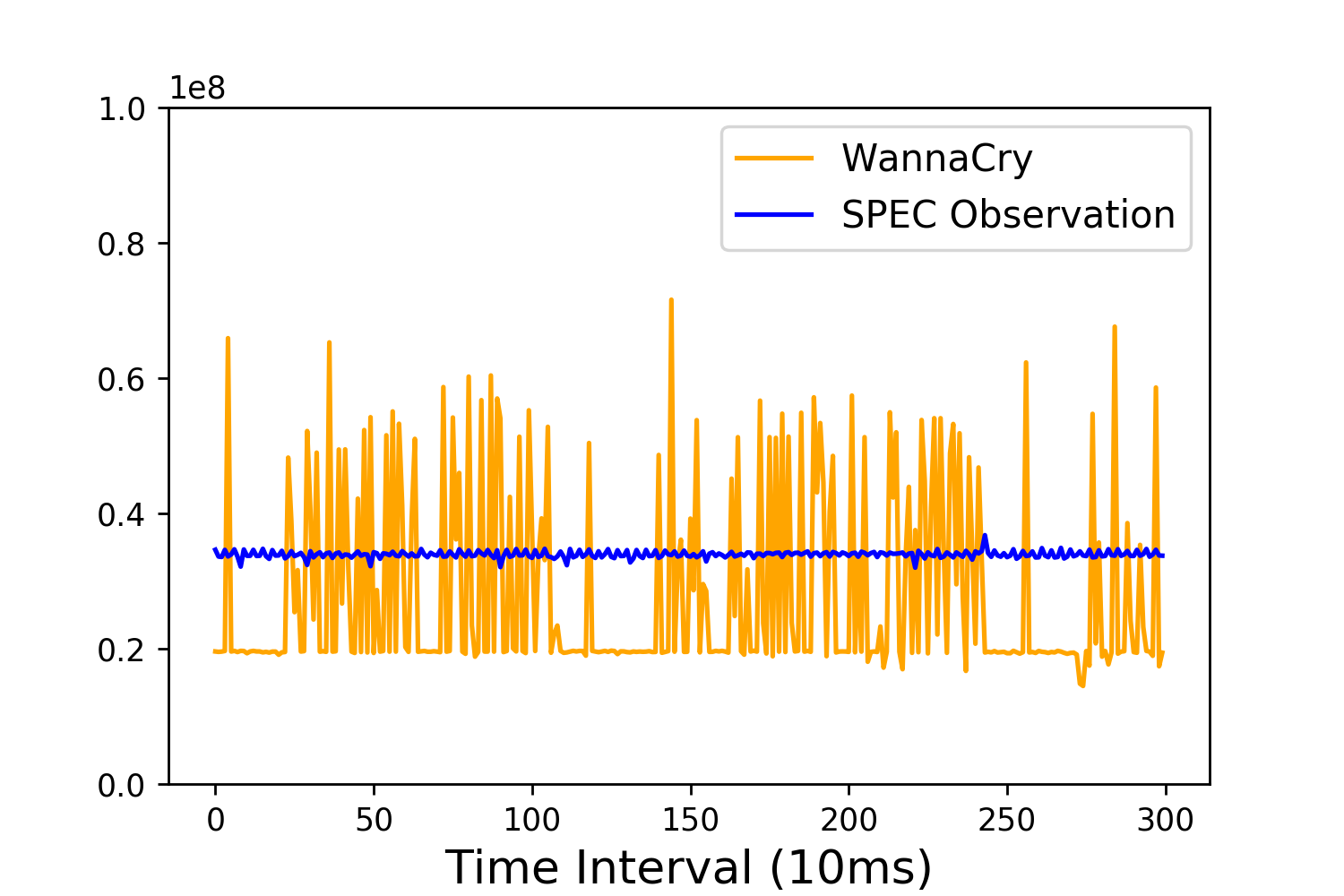}
		\label{fig:com_ins}}\vspace*{-0.5cm}%\vspace{-0.65cm}
	\caption{\small Comparison of the Effects on Performance Event Counters from HPCs 
		in presence of Wannacry Ransomware and SPEC Benchmark Programs\label{fig:perf_dev_com}}\vspace*{-0.6cm}
\end{figure}

\begin{figure}[!b]
		\centering
		\includegraphics[width=0.46\linewidth]{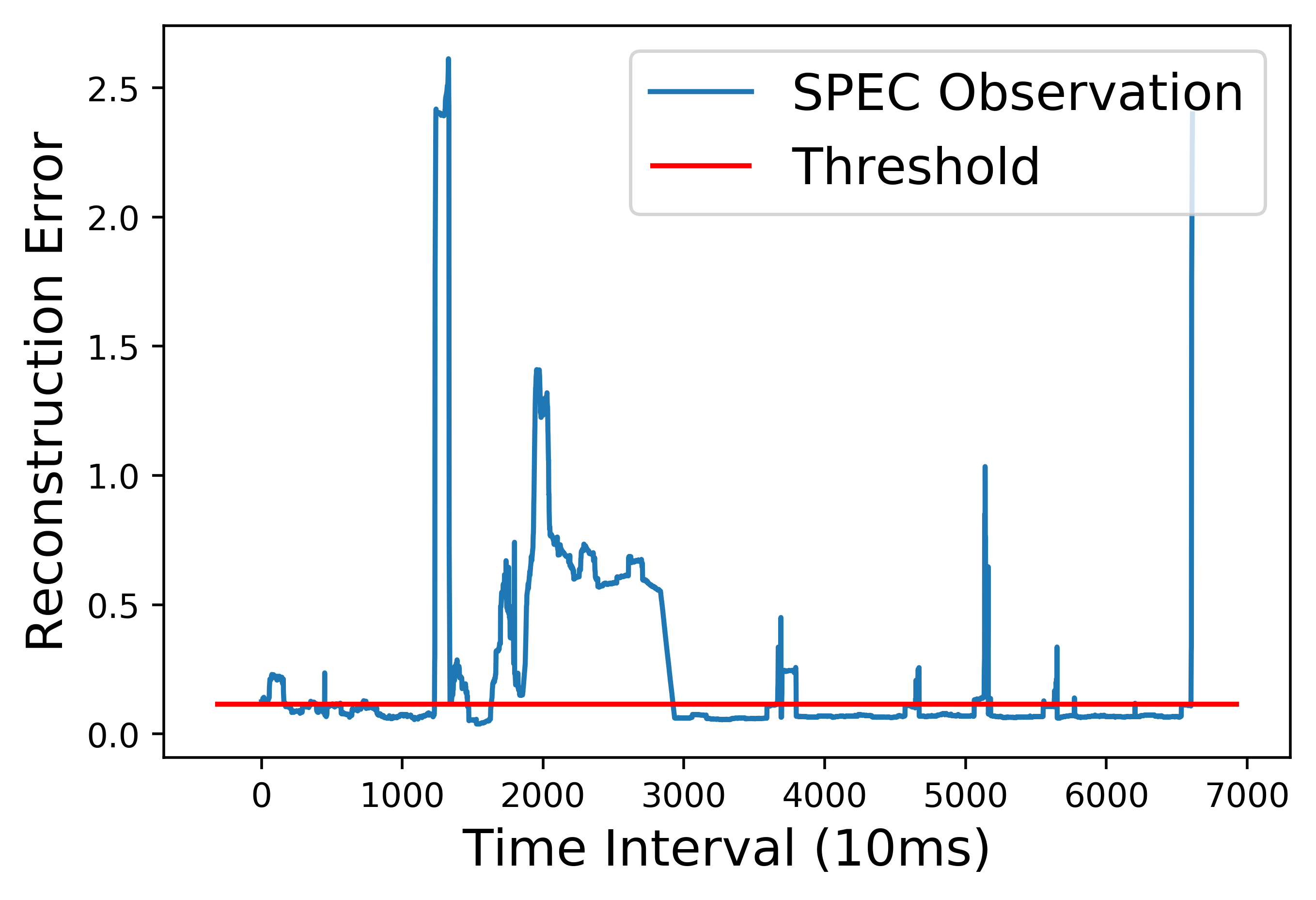}\vspace*{-0.5cm}
		\caption{\small Sequence of Reconstruction Errors for SPEC Benchmark Programs}\vspace*{-0.4cm}
		\label{fig:rec_spec}
	\end{figure}

	In order to test the robustness of our detection scheme, we incorporate 
	an analysis in presence of SPEC2006 server and multimedia benchmarks. 
	We consider the Gshare predictor implementation 
	as provided in ($https://www.jilp.org/jwac-2/cbp3\_framework\_\\instructions.html$) and observe the 
	HPC sampling counts from a background process exactly like our previous setting. Figure~\ref{fig:perf_dev_com}
    presents the variation of different hardware events in presence of both SPEC benchmarks and WannaCry
    ransomware. We can observe that the execution behavior for both the programs are significantly different
    from the normal observations. Thus, the sequences of data for the SPEC programs may also create considerable
    reconstruction errors. In Figure~\ref{fig:rec_spec}, it clearly shows that the reconstruction error for the sequences
	in presence of the SPEC benchmark programs is above the predetermined threshold at the first
	window itself. Though the error is very close to the threshold, this essentially raises an alarm to
	RAPPER that this benchmark program is a potential malicious	program which deviates in an extent
	from the normal system behavior.
	But surely in this case, it is a false alarm, since the benchmark is composed of 
	server and multimedia benchmarks and can be considered as the representative of the 
	high computational processes which may deviate highly to the normally running processes in a system. 

\begin{figure*}[!h]
	\centering
	\small %\vspace{-0.55cm}
	\subfigure[\textbf{\small FFT of branch Instructions}]{
		\includegraphics[width=0.26\textwidth]{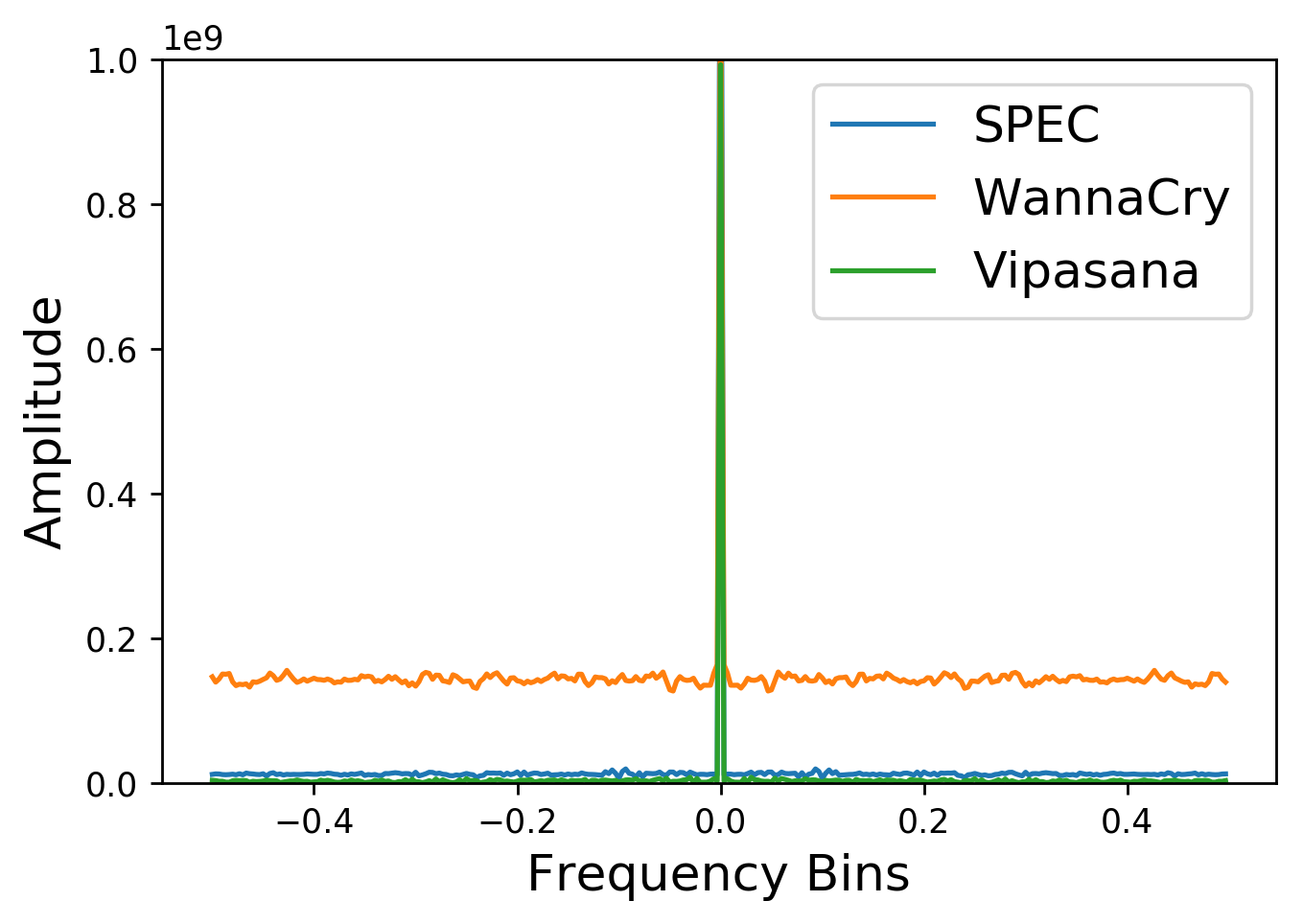}
		\label{fig:fft_branch}}
	\quad
	\subfigure[\textbf{\small FFT of branch mispredictions}]{
		\includegraphics[width=0.26\textwidth]{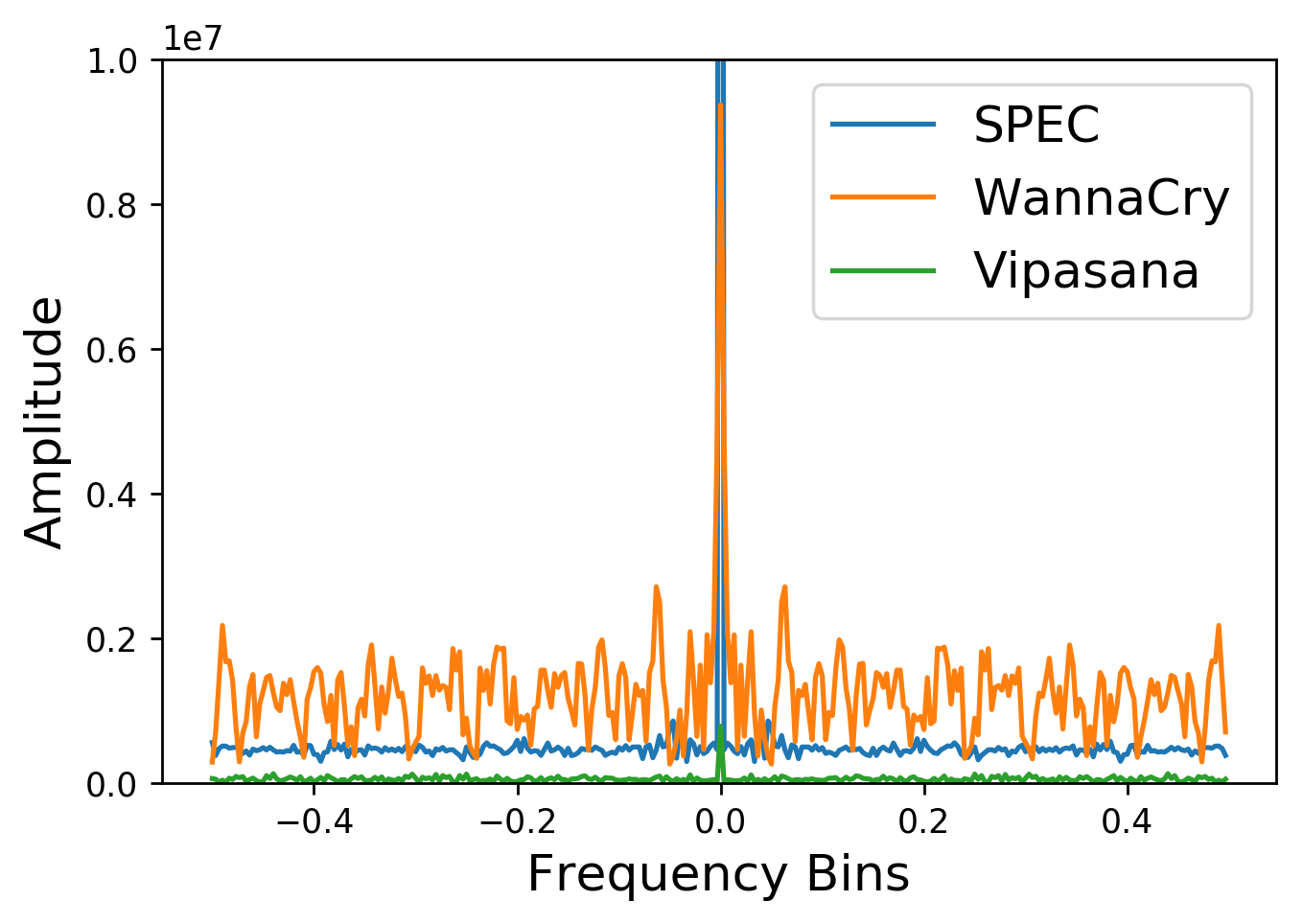}
		\label{fig:fft_br_miss}}%\vspace{-0.65cm}
	\quad
	\subfigure[\textbf{\small FFT of cache misses}]{
		\includegraphics[width=0.26\textwidth]{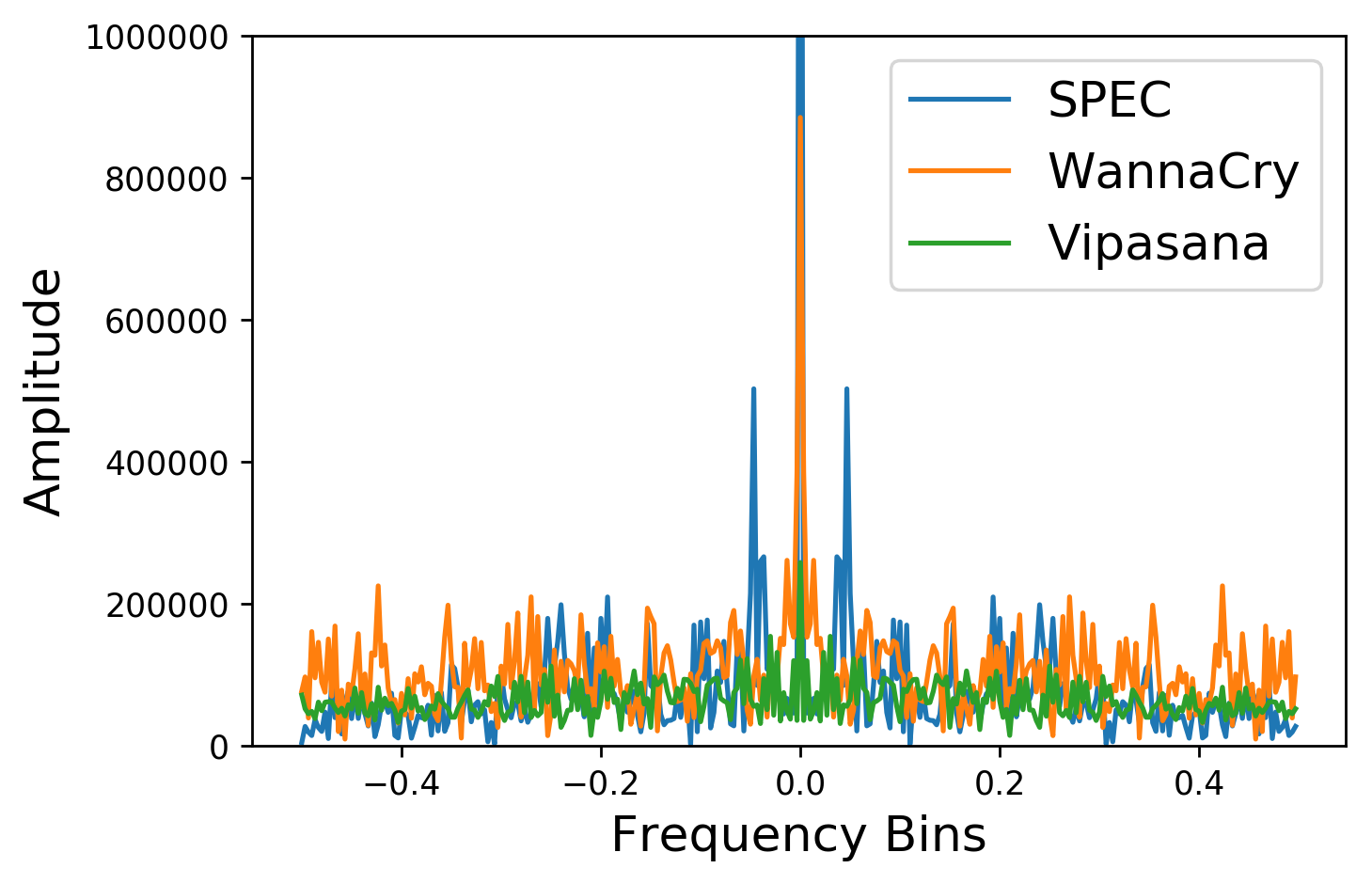}
		\label{fig:fft_ca_miss}}%\vspace{-0.65cm}
	\quad
	\subfigure[\textbf{\small FFT of Cache references}]{
		\includegraphics[width=0.26\textwidth]{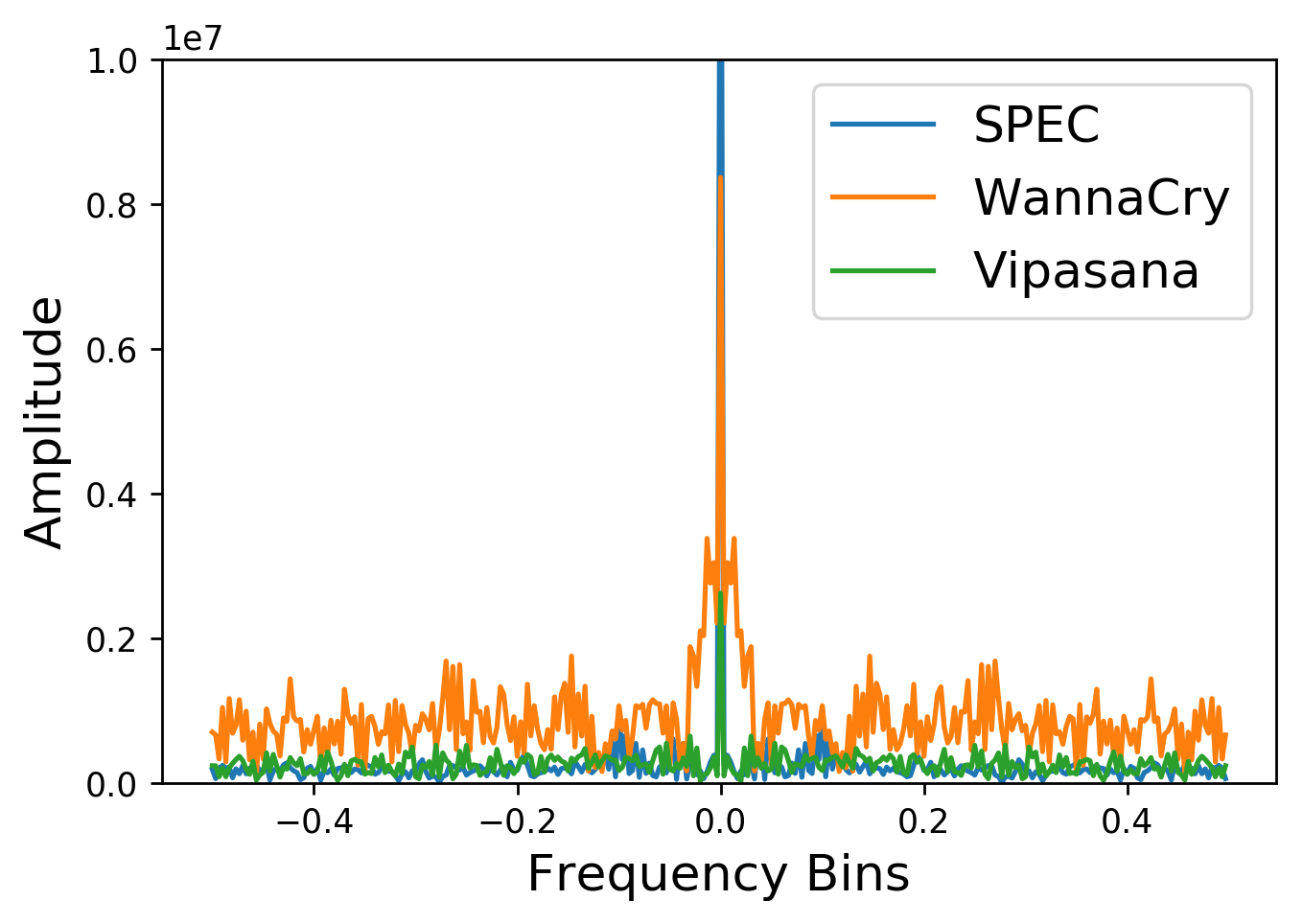}
		\label{fig:fft_cache}}%\vspace{-0.65cm}
	\quad
	\subfigure[\textbf{\small FFT of Instructions}]{
		\includegraphics[width=0.26\textwidth]{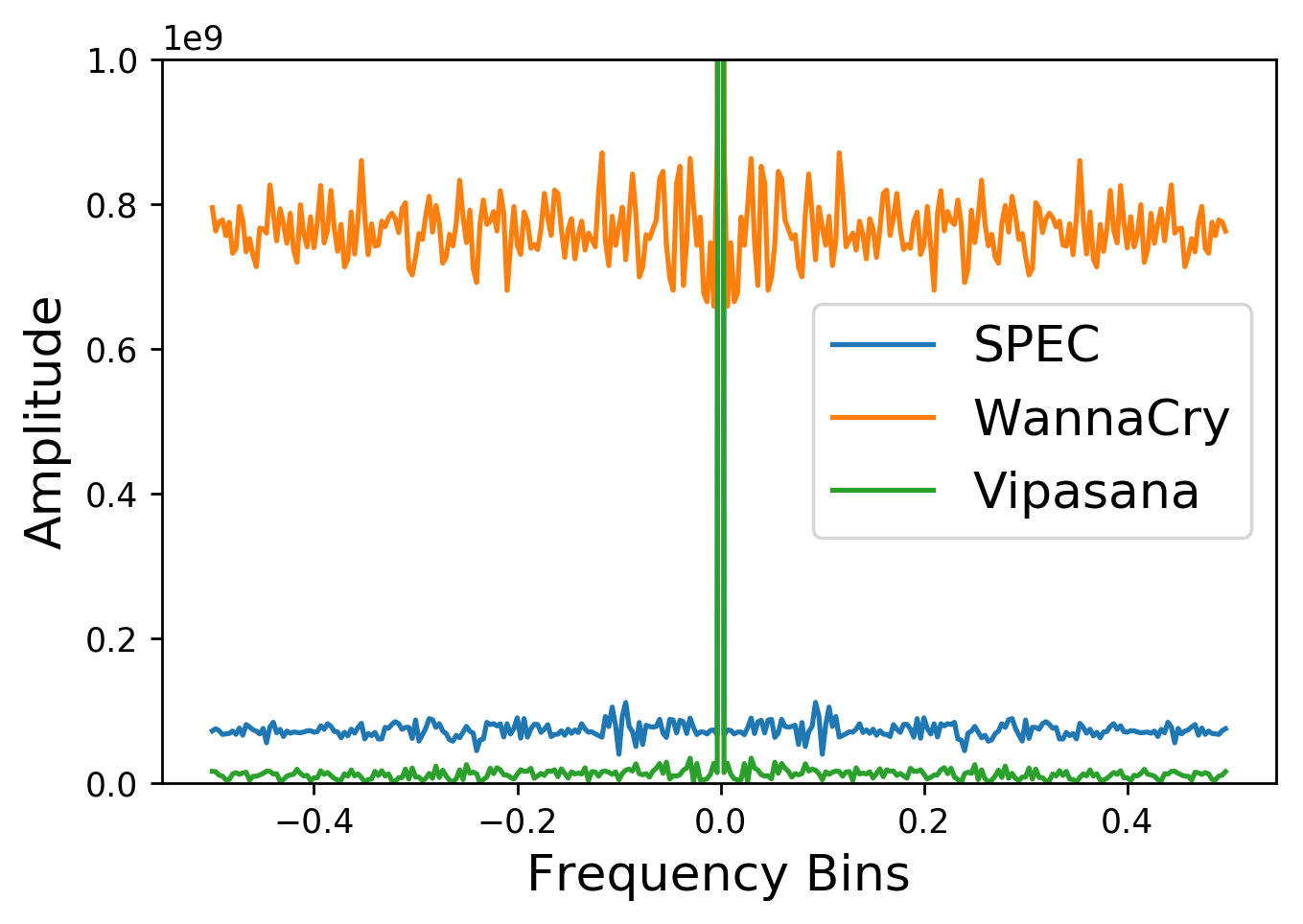}
		\label{fig:fft_ins}}\vspace*{-0.5cm}%\vspace{-0.65cm}
	\caption{\small Variation of Amplitude in frequency domain of the performance counters from HPCs in the presence of SPEC observation and Wannacry Ransomware\label{fig:fft_dev}}\vspace*{-0.4cm}
\end{figure*}
In the next subsection, we perform a transformation from the time domain to the frequency domain to differentiate actual malicious processes from false positives.

%\vspace{-0.2cm}
\subsection{Introducing Fast Fourier Transformation}
In the second phase of detection using RAPPER, we transform the traces from the 
time domain to the frequency domain using the Fast Fourier Transformation(FFT). 
FFT is the most efficient way to implement the Discrete Fourier Transformation. 
The primary reason to convert the analysis from the time domain to the frequency domain is to understand the repetitive pattern of the traces. The ransomware executable runs encryption repeatedly on multiple files thus it repeats a same set of operations of opening a file, encrypting and closing the file followed by deleting it for multiple files one after another. 
The transformation is illustrated in Figure~\ref{fig:fft_dev}, which typically indicates that the amplitude for each frequency bins are constantly higher for the ransomware in contrary to the SPEC benchmark.

We have applied FFT on the time domain values for different hardware events as mentioned in Section~\ref{sec:hpc}, 
to obtain the frequency domain values. Figure~\ref{fig:fft_dev} presents the FFT plots for the normal system measurements in \textcolor{blue}{blue} lines, along with the SPEC Observations
in \textcolor{green}{green} lines and 
WannaCry Ransomware in \textcolor{orange}{orange} lines for different hardware events. 
Figure~\ref{fig:fft_dev} shows that for most of the hardware events (apart from the cache misses), the FFT plot behavior of the SPEC benchmark overlaps exactly with the FFT of the normal system behavior. 
Also, it is quite clear from the Figure~\ref{fig:fft_branch}, Figure~\ref{fig:fft_br_miss},
Figure~\ref{fig:fft_cache}, and Figure~\ref{fig:fft_ins} that the amplitude of almost all the frequency bins are
higher for WannaCry than the SPEC observation, which is eminent as the WannaCry program repeatedly encrypt
multiple files.

\begin{figure}[!h]
		\centering
		\small %\vspace{-0.55cm}
		\subfigure[\textbf{\small SPEC Benchmark}]{
			\includegraphics[width=0.45\linewidth]{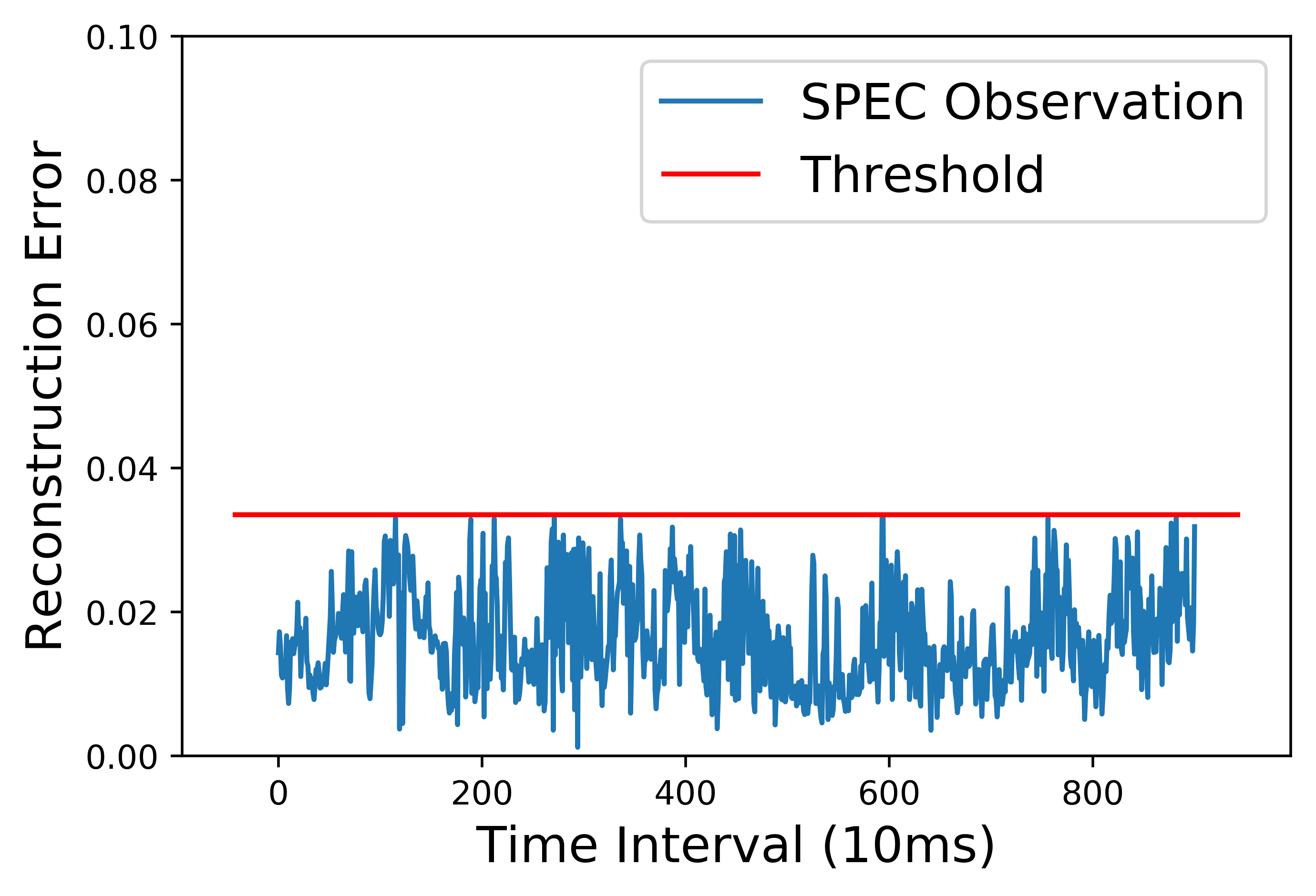}
			\label{fig:rec_wannacry_2}}
		\subfigure[\textbf{\small WannaCry}]{
			\includegraphics[width=0.45\linewidth]{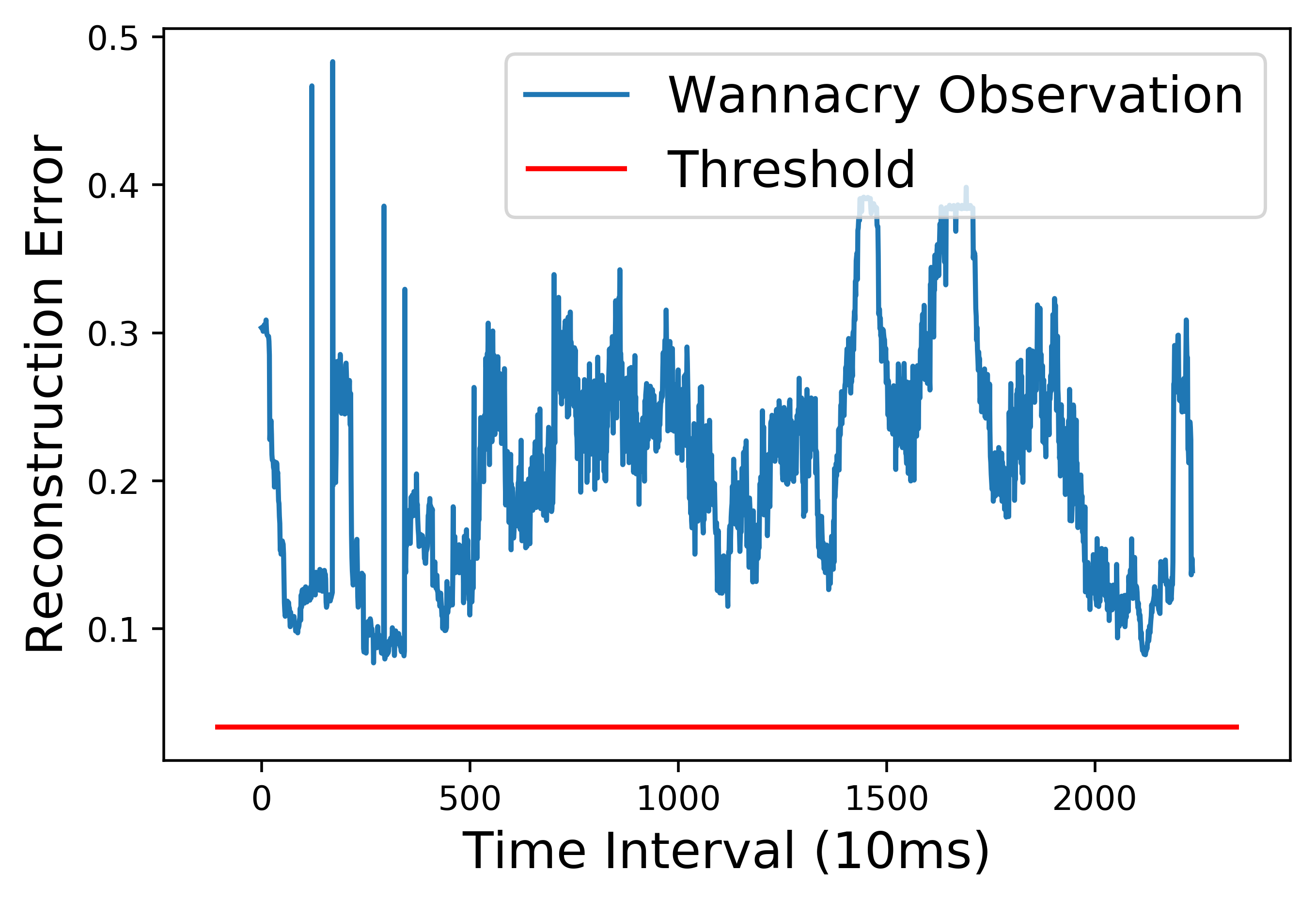}
			\label{fig:rec_vipasana_2}}\vspace*{-0.5cm}%\vspace{-0.65cm}
		\caption{\small Sequence of Reconstruction Errors for SPEC Benchmark (a) and WannaCry (b)\label{fig:seq_rec_error_2}}\vspace*{-0.4cm}
	\end{figure}

The detection of these variations of amplitudes for different frequency bins can again be considered
as a time-series data, and an LSTM based autoencoder, as discussed before, can be used to detect the
anomaly. The amplitudes for SPEC benchmark programs are very close to that of regular observations 
for most of the hardware events. Thus, modeling the FFT data for regular sequences using an autoencoder 
will result in reconstruction errors close to the threshold (say $\mathcal{R}^{\prime}_t$) for SPEC benchmarks, and
the error will be much higher in case of ransomwares because of the repeated encryptions.
We modeled an autoencoder as mentioned before and the calculated the threshold $\mathcal{R}^{\prime}_t$ 
to be $0.033$. Fig~\ref{fig:seq_rec_error_2} presents the sequence of reconstruction errors for both SPEC and WannaCry programs
and we can verify that the reconstruction errors of the SPEC programs always lies below the threshold and thus discarded from 
false positives, whereas the reconstruction error of the WannaCry program always remains higher to the threshold.

%\vspace{-0.2cm}
\section{Architecture of RAPPER}
In this section, we present an overview of the architecture of proposed detection methodology - RAPPER. The basic
diagram of the system is shown in Figure~\ref{fig:architecture}. All the experimentation for this study have been performed in a sandbox environment, such that the ransomwares do not affect the actual filesystem.

The architecture contains four modules (\emph{Watchdog Program}, \emph{Autoencoder\_1}, \emph{FFT Converter}, and \emph{Autoencoder\_2}). The detection methodology works in two phases, namely \emph{Offline Phase} and \emph{Online Phase}.
The functioning of each of the module in both the phases are decribed below:
\vspace{-0.2cm}
\subsection{Offline Phase}
In the offline phase, the detection methodology is trained with the normal behavior of the sandbox environment, such that any unusual activity of a ransomware is properly detected in real-time scenario. The functioning of each of the modules in this phase are described below.
\begin{enumerate}{\small
\item \emph{Watchdog Program}: Monitors the HPCs of the Sandbox Environment continuously and forwards a window of data (calculated as described before) to the Autoencoder\_1 and the FFT Converter in parallel.
\item \emph{Autoencoder\_1}: Collects all the data forwarded by the watchdog program and an autoencoder is trained with the dataset as mentioned in Section~\ref{sec:learn_auto}.  
\item \emph{FFT Converter}: Converts the Fast Fourier Transformation of each window forwarded by the watchdog program and passes the results to the Autoencoder\_2.
\item \emph{Autoencoder\_2}: Collects all the data passed by the FFT Converter and trains another autoencoder based on the FFT dataset.}
\end{enumerate}
%\vspace{-0.2cm}

\subsection{Online Phase}
In the online phase, the detection module is deployed in the sandbox system for real-time monitoring to detect 
ransomwares. The functioning of each modules in this phase are discussed below.
\begin{enumerate}{\small
\item \emph{Watchdog Program}: Monitors the sandbox system as performed in training phase, and forwards the data
to the Autoencoder\_1 module. In this phase, watchdog program does not forward data to the FFT converter. This
helps us to monitor the system with lower computational cost.
\item \emph{Autoencoder\_1}: Receives sequence window at each time interval from the watchdog program and calculates
the reconstruction error of the sequence. If the error is higher than the predefined threshold $\mathcal{R}_t$, it
sends a signal to the watchdog program to transmit the same window to the FFT Converter.
\item \emph{FFT Converter}: Receives a sequence window from the watchdog module, converts the data into frequency domain,
and forwards the transformed data to the Autoencoder\_2 as done in training phase, but with a condition imposed by the
Autoencoder\_1 module.
\item \emph{Autoencoder\_2}: Calculates the reconstruction error of the received FFT data, and based on the predefined
threshold, $\mathcal{R}^{\prime}_t$, sends a warning whether the sequence belongs to a ransomware or not.}
\end{enumerate}

\begin{figure}[!t]
  \centering
   \includegraphics[width=\linewidth, height = 6cm]{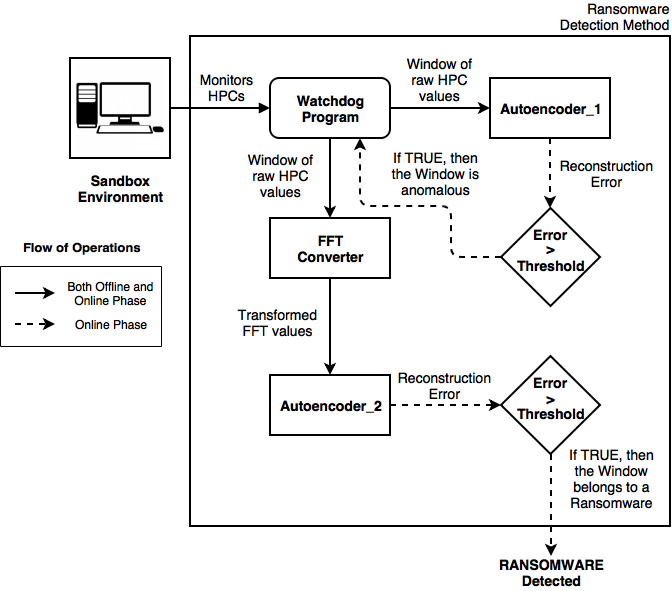}\vspace*{-0.4cm}
  \caption{Detection Methodology of RAPPER\label{fig:architecture}}\vspace*{-0.4cm}
\end{figure}

%\vspace{-0.2cm}
\subsection{Evaluating the performance of RAPPER}\label{sec:results}
We performed all the experiments in a sandbox system having specification \texttt{Intel Core i3 M350} running Ubuntu 16.04 with 4.10.0-38-generic kernel.
We used popular open source python based neural network library \texttt{Keras}~\cite{keras} for the implementation
of both the autoencoder. The architecture used to model the autoencoders are mentioned in Table~\ref{table:architec}. The Dropout Layer is added to
regularize the neural network and prevent from overfitting (Dropout
is a technique where a set of randomly selected neurons are ignored during training).

\begin{figure}[!t]
	\centering
	\small %\vspace{-0.55cm}
	\subfigure[]{
		\includegraphics[width=0.45\linewidth]{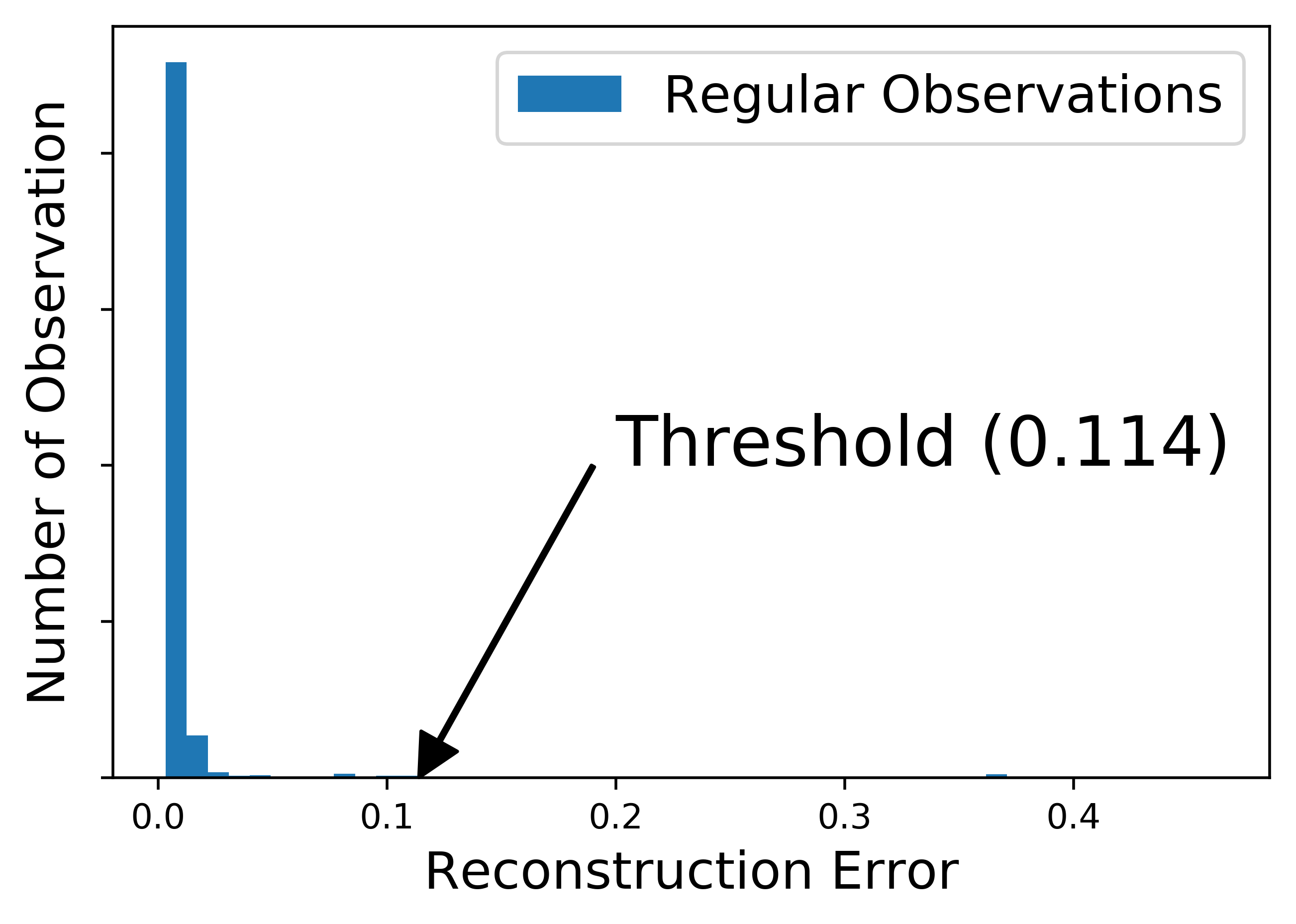}
		\label{fig:threshold1}}
	\subfigure[]{
		\includegraphics[width=0.45\linewidth]{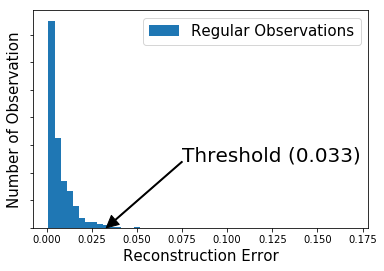}
		\label{fig:threshold2}}\vspace*{-0.5cm}%\vspace{-0.65cm}
	\caption{Distribution of reconstruction errors for (a) Autoencoder\_1 (b) Autoencoder\_2\label{fig:threshold}}\vspace*{-0.4cm}
\end{figure}

The distribution of reconstruction errors for regular observations produced by both the
autoencoders are shown in Figure~\ref{fig:threshold}. The thresholds thus calculated using 
Equation~\ref{eq:three_sigma} are $0.114$ and $0.033$ respectively. 

The FFT converter usually takes $0.0003$ milliseconds to convert a sequence within a window into frequency domain.
The model building times for Autoencoder\_1 and Autoencoder\_2 are on average $10$ and $14$ minutes respectively. Testing time to calculate whether a single window is an anomaly or not
is $1.321$ milliseconds for Autoencoder\_1 and $1.699$ milliseconds for Autoencoder\_2 respectively. As shown in the Architecture of RAPPER in Figure~\ref{fig:architecture}, the testing of a regular observation only passes through 
the Autoencoder\_1, thereby taking only $1.699$ milliseconds, and an anomalous observation passes through all the three modules: Autoencoder\_1, FFT Converter, and Autoencoder\_2, thereby taking $1.321+0.0003+1.699 = 3.0203$ milliseconds to be detected. However, in both the cases, the detection time is less that the sampling interval, which is $10$ milliseconds. Hence, the
detection is performed seamlessly, without the need of any storage buffer, as a new window of data will be created after 10 milliseconds.

Without loss of generality, we measured the performance of RAPPER on most recent WannaCry ransomware. As shown in Section~\ref{sec:threshold}, the WannaCry is detected as an anomaly at the $245^{th}$ window and instantly detected as
ransomware at the same time because it's reconstruction error is always higher than the threshold of Autoencoder\_2. Hence,
the total time taken to detect WannaCry is equal to (Time taken to generate the first window) + $244$ * (time interval for each 
sample) + (Autoencoder\_1 testing time) + (Time for single FFT Conversion) + (Autoencoder\_2 testing time) 
$=1000+244*10+1.321+0.0003+1.699$ millisecond $=3443.0203$ milliseconds. Thus, WannaCry is detected by RAPPER in approximately $3.443$ seconds. As a sample run with RAPPER, out of 10000 files of approximately 21 bytes each, when the detection stops the execution, 68 files are encrypted. It maybe noted that, the size of a typical file is much larger than 21 bytes, and hence, a lesser number of files will be encrypted.

%\vspace{-0.2cm}
\section{File Recovery and Conclusions}
RAPPER is thus capable of detecting the presence of ransomwares 
fast, as we show for the case of WannaCry within a time 
of approximately 4 sec from its launch. Depending on the latency, 
the malware can encrypt few files (say $n$). We conclude with a 
suggested approach on data retrieval. 
A practical solution would be to take backups of 
the $n$-recently opened files. After 
the lapse of the time quantum required to encrypt these files, 
we delete the copies if no ransomware alarm is raised by RAPPER. 
This minimizes the storage necessary for the backup files. 
To further ensure that the backup files are not encrypted 
we perform locking operation, like in linux using {\tt mlock}. 

In this paper, thus we explored the effect of ransomware on normal system behaviors. We take the aid of the Artificial Neural Network to detect the presence of ransomwares using a two-step detection framework. The entire detection procedure does not need any template of the malicious process beforehand. Instead it thrives on an anomaly detection procedure to detect the infectious ransomwares in as less as 4 seconds with almost zero false positives, using a frequency analysis. 

We also explored the opportunity of applying 
side channel techniques to recover the secret key used to 
encrypt the files from the perf statistics. Wlog. we 
found for ransomwares like WannaCry, each file 
is encrypted using AES-128 CBC (Cipher Block Chaining) with a randomly generated distinct key.  These keys are in turn encrypted using an infection 
specific RSA public key
and stored in the memory. 
It would be indeed a challenging exercise to recover the 
AES key by targeting the AES CBC operation. 
However, we leave that as a future scope of work.

\begin{table}[!t]
\centering
{\footnotesize
\caption{Model Architecture for Autoencoders} \vspace*{-0.2cm}
\label{table:architec}
\begin{tabular}{|c|c|c|c|}
\hline
\textbf{\begin{tabular}[c]{@{}c@{}}Layer\\ Number\end{tabular}} & \textbf{\begin{tabular}[c]{@{}c@{}}Layer\\ Type\end{tabular}} & \textbf{\begin{tabular}[c]{@{}c@{}}Input\\ Shape\end{tabular}} & \textbf{\begin{tabular}[c]{@{}c@{}}Output\\ Shape\end{tabular}} \\ \hline
\multicolumn{4}{|c|}{\textbf{Autoencoder\_1}} \\ \hline
1 & LSTM & (None, 100, 5) & (None, 100, 32) \\ \hline
2 & Dropout & (None, 100, 32) & (None, 100, 32) \\ \hline
3 & LSTM & (None, 100, 32) & (None, 100, 5) \\ \hline
\multicolumn{4}{|c|}{\textbf{Autoencoder\_2}} \\ \hline
1 & LSTM & (None, 100, 5) & (None, 100, 64) \\ \hline
2 & Dropout & (None, 100, 64) & (None, 100, 64) \\ \hline
3 & LSTM & (None, 100, 64) & (None, 100, 5) \\ \hline
\end{tabular}}\vspace*{-0.4cm}
\end{table}

%\begin{scriptsize}
\bibliographystyle{ACM-Reference-Format}
\bibliography{refs}
%\end{scriptsize}
\end{document}